\documentclass[12pt]{article}

\newenvironment{wileykeywords}{\textsf{Keywords:}\hspace{\stretch{1}}}{\hspace{\stretch{1}}\rule{1ex}{1ex}}

\usepackage{amsmath,amssymb}
\usepackage{graphicx}
\usepackage{color}
\usepackage{dcolumn}
\usepackage{bm}
\usepackage[numbers,super,comma,sort&compress]{natbib}

\definecolor{background-color}{gray}{0.98}

\usepackage{url}

\usepackage{hyperref}
\usepackage{array}
\usepackage{longtable}
\usepackage{diagbox}
\usepackage{multirow}
\usepackage{siunitx}
\usepackage[table]{xcolor}

\title{Machine learning-assisted search for novel coagulants: when machine learning can be efficient even if data availability is low}
\author{%
Andrij Rovenchak\footnote{SoftServe, Inc., 2d Sadova St., 79021 Lviv, Ukraine}~\footnote{Professor Ivan Vakarchuk Department for Theoretical Physics, Ivan Franko National University of Lviv, 12 Drahomanov St., 79005 Lviv, Ukraine
},
Maksym Druchok\footnotemark[1]~\footnote{Institute for Condensed Matter Physics, 1 Svientsitskii St., 79011 Lviv, Ukraine
}}

\begin{document}

\maketitle

\begin{abstract}
Design of new drugs is a challenging process: a candidate molecule should satisfy multiple conditions to act properly and make the least side-effect -- perfect candidates selectively attach to and influence only targets, leaving off-targets intact.
The amount of experimental data about various properties of molecules constantly grows, promoting data-driven approaches. 
However, the applicability of typical predictive machine learning techniques can be substantially limited by a lack of experimental data about a particular target.
For example, there are many known Thrombin inhibitors (acting as anticoagulants), but a very limited number of known Protein C inhibitors (coagulants).
In this study, we present our approach to suggest new inhibitor candidates by building an effective representation of chemical space.
For this aim, we developed a deep learning model -- autoencoder, trained on a large set of molecules in the SMILES format to map the chemical space.
Further, we applied different sampling strategies to generate novel coagulant candidates.
Symmetrically, we tested our approach on anticoagulant candidates, where we were able to predict their inhibition towards Thrombin.
We also compare our approach with MegaMolBART -- another deep learning generative model, but exploiting similar principles of navigation in a chemical space.
\end{abstract}

\begin{wileykeywords}
molecular design, 
machine learning,
coagulants, 
anticoagulants.
\end{wileykeywords}

\clearpage

\begin{figure}[h]
\centering
\colorbox{background-color}{
\fbox{
\begin{minipage}{1.0\textwidth}
\includegraphics[width=50mm,height=50mm]{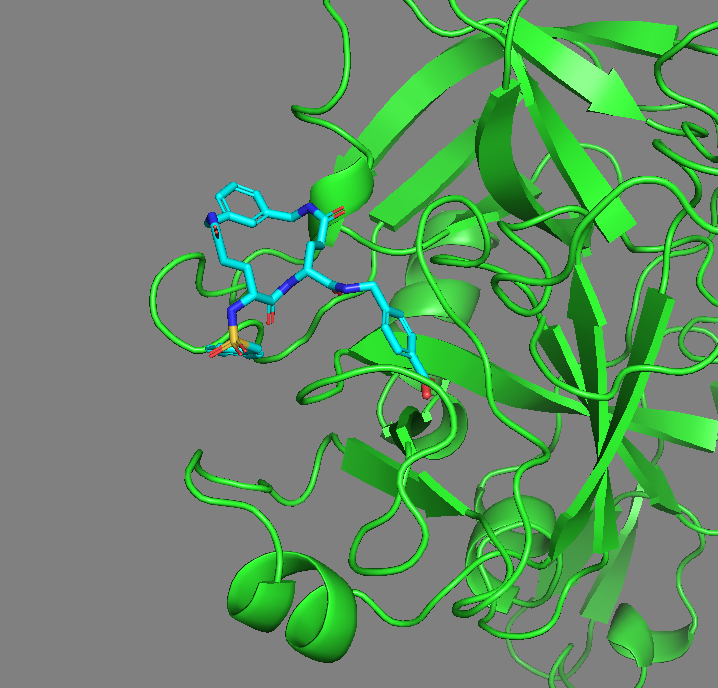} 

This study employs machine learning to generate new drugs, emphasizing cases with low data availability. 
Focusing on coagulants, underrepresented in databases, our approach generates molecular encodings based on the assumption that similar structures share properties.
Strategies tested on anticoagulants are applied to discover novel coagulant candidates, navigating the encoding space.

\end{minipage}}}

\end{figure}

  \makeatletter
  \renewcommand\@biblabel[1]{#1.}
  \makeatother

\bibliographystyle{apsrev}

\renewcommand{\baselinestretch}{1.5}
\normalsize

\clearpage

\section*{\sffamily \Large INTRODUCTION}\label{sec:Intro}

This study is intended to explore a path of suggesting novel compounds by machine learning~(ML) techniques with a specific case of limited data availability.
In particular, we aim to find new coagulants, which are weakly represented in specialized databases, as much as, for example, anticoagulants.
But before we dive into coagulants specifics, we would like to give a broader overview of current status of ML applications in drug design.
A drug discovery process begins with localization of a disease cause, drafting of a list of drug candidates, and screening them \textit{in silico}.
During this screening, multiple properties of compounds are assessed to filter out least potent candidates and shrink the list before it gets to \textit{in vitro} tests.
Obviously, drafting and filtering influence the effectiveness of the downstream drug design stages quite substantially, as passing too many weak candidates will waste time and resources.
Since structure and chemical composition of compounds define their properties, the task of establishing Quantitative Structure-Activity/Property Relationship~(QSAR) relates structure of compounds with their physical/chemical properties, e.g., solubility in water or organic solvents, melting temperature, solvation energies, etc.~\cite{Lusci2013,Pyzer-Knapp2015,Schneider2018}

During the last decade, we witness a wave of ML-powered approaches in biochemical domain (see, for example, Refs.~\citenum{Alipanahi2015,Mayr2016,Sharma2017,Stahl2018,Zagotto2021,Pandiyan2022,Hagg2023}).
There are a few favoring factors here: (i) successes of ML in other areas, like computer vision, autonomous driving, natural language processing, (ii) growing availability of data, (iii) complex problems can be solved in a data-driven manner (in particular, artificial neural networks are often referred as ``universal approximators'').
To our best knowledge, first attempts to apply ML in drug design date back to the early and mid-1990s,~\cite{Bolis1991,Jain1994,Clark1995} while they can be traced even further back to applications in QSAR~\cite{Hansch1964,King1993,Grover2000,Gad2014} and, ultimately, to the seminal idea of Crum Brown and Fraser~\cite{CrumBrown1868} that biological properties can be expressed as a function of the chemical structure.
A structure-property relation is also actively used in biochemistry to predict drug-likeliness and ADMET (Absorption, Distribution, Metabolism, Excretion, Toxicity) scores.
A cornerstone problem here is the prediction of binding affinities between drug candidates and biomolecules.
The complexity arises from a huge combinatorial space of structural variations of organic molecules.
Hence, solving the problem by conventional computer simulation approaches is time- and resource-consuming, while lead selection requires screening of hundreds of thousands of candidate molecules.~\cite{Steinhauser2009}
One of the pioneering studies with a specific focus on binding affinity prediction by ML is reported in Ref.~\citenum{King1995} where the performances of neural networks, k-Nearest Neighbors, and Decision Trees models are compared on the set of about 200 ligands and two target receptors.
The list of ligand descriptors there includes molecular size, flexibility, polarity, polarizability, numbers of donors/acceptors, etc.
In a study by Jorissen \textit{et~al.},~\cite{Jorissen2005} Support Vector Machine is used to predict the binding affinity of compounds.
A number of studies from different research groups~\cite{Pahikkala2014,He2017,Ozturk2018,Ozturk2019,Shin2019,Nguyen2020,Shim2021} are oriented on KIBA~\cite{Tang2014} and Davis~\cite{Davis2011} kinase datasets.
These studies use different ML approaches to build predictive models for the binding affinity between ligands and kinases.
A study by Kundu~\textit{et al.}~\cite{Kundu2018} compares a performance of multiple ML methods on affinity prediction on the PDBbind dataset.~\cite{PDBbind:www,Wang2004,Liu2017}
The DGraphDTA approach~\cite{Jiang2020} builds graphs for ligands and proteins to feed them to a graph neural network, which is trained to predict binding affinities.
$K_\textrm{deep}$~\cite{Jimenez2018} and DeepAtom~\cite{YLi2019} are three-dimensional convolutional neural networks for binding affinity prediction.

It is also worth mentioning a series of studies utilizing an ensembling of ML models.~\cite{Chen2019,Kwon2020,Schneider2020,Druchok2021}
The strategy behind such an ensembling is based on the assumption that different ML techniques have their own peculiarities, thus, ensembling helps cross-compensate them and provide more robust predictions.

Training of binding affinity predictors can be done with respect to either single or multiple receptors.
In the first case, a receptor is fixed and not considered as an input to the ML model, hence, only ligands (with known affinities towards this receptor) take part in the training.~\cite{Druchok2021}
This approach is more convenient in the case of high data availability, but, obviously, such predictions are valid towards only a single receptor.
Multi-receptor models are trained on multiple ligands and multiple receptors.~\cite{Nikolaienko2022}
These models benefit from knowledge of affinities of different ligand-receptor pairs and can potentially generalize to previously unseen combinations, however, prediction errors in this case remain relatively high, prompting for more experimental data to be collected.~\cite{Nikolaienko2022,Zhu2022}

Alternatively, in the case of low data availability about known inhibitors of a particular receptor, one can rely on a hypothesis that structurally similar compounds possess similar inhibitory action.
Hence, new inhibitors can be drafted via generation of similar compounds in the vicinity of known inhibitors and the problem reformulates to ``mapping of a chemical space'', which, in particular, can be done with autoencoders.
The idea of autoencoders, a type of generative neural network model, originates from the early and mid-2010s.~\cite{Kingma2014,Rezende2014}
Its application ranges from language models~\cite{Bowman2016} and image processing,~\cite{Zhou2018} medical imagery,~\cite{Zhao2019,Ternes2022} to design of chemical compositions and structures~\cite{Vasylenko2021} or even social sciences problems.~\cite{Huang2019}
One of the most recognized studies applying variational autoencoder in drug discovery was reported in 2018: authors used SMILES (Simplified Molecular Input Line Entry System)~\cite{Weininger1988,Weininger1989,SMILES:www} strings as an input and encoded them into vectors in a latent space of continuous variables.~\cite{GomezBombarelli2018}
To mention a few other examples: different variations of the autoencoder concept were implemented in the last years.~\cite{Lim2018,Joo2020,Druchok2021b,Zhang2022,He2022}

After such an overview of ML applications to biochemical domain, we would like to proceed with coagulant-specific considerations.
With this respect, one needs to mention Protein C -- a natural anticoagulant protein produced in liver and found in bloodstream.
It is synthesized as an inactive precursor requiring activation to exert its anticoagulant effect.
The Protein C pathway is initiated when there is an activation of the coagulation cascade, typically in response to injury or damage to blood vessels.
This activation ignites production of Thrombin, a key enzyme in blood clot formation.
Thrombin not only promotes clot formation but also has the ability to activate Protein C.~\cite{Kisiel1979}
Once activated, Protein C binds to a co-factor called Protein S.
Activated Protein C and Protein S together form an anticoagulant complex that inactivates two important clotting factors, Factors Va and VIIIa (co-factors in activation of Factor X and Prothrombin), limiting the formation of blood clots and preventing them from growing too large and causing excessive clotting.
As the bottom line, by inhibiting Protein C (reducing anticoagulant function), one can influence clot formation and, thus, enhance coagulation.
A more detailed description of the Protein C pathway can be found in Refs.~\citenum{Esmon2003,Dahlback2005}: there are multiple steps involved in the balance between coagulant and anticoagulant functions of an organism, but for simplicity we limit the subject of this study to Protein C only, as presented approach can be applied to other proteins as well.

The remaining part of the paper is organized as follows. 
Methodological background of the approach is described in the \hyperref[sec:Methods]{Methods} section, followed by the \hyperref[sec:Data]{Data} section.
Then, the study outcomes are presented in the \hyperref[sec:Results]{Results} section, 
while the \hyperref[sec:Conclusions]{Conclusions} section summarizes the study.

\section*{\sffamily \Large METHODS}
\label{sec:Methods}

\subsection*{\sffamily \large General outline}

Usually, research groups, specializing in ML, tackle inhibitor design-related problems by combining generative and predictive approaches.
Generative approaches iteratively suggest candidates, while predictive ones (trained on known examples) assess the candidate quality and provide feedback to adjust the generation course.
This can be a scenario for anticoagulants (Thrombin inhibitors) search, however, there are not many known coagulants to train a confident predictive model.
Under conditions of little knowledge about inhibitors, we will rely on the hypothesis that structural similarity yields functional similarity, mentioned in the Introduction.

First, we build our autoencoder to map a chemical space of small organic molecules.
The autoencoder has two parts -- encoder and decoder.
Both encoder and decoder consist of multiple neural layers, where encoder's purpose is to reduce the input size of a sample to some small bottleneck, while the decoder revives the dimensions back to the original size.
The autoencoder is considered trained when encoder input vectors coincide with decoder output ones.
Data passing through the bottleneck is called an embedding vector, thus, the function of the encoder and decoder is to compress and decompress the information about the data sample.
All embeddings constitute a hyperspace of chemical compounds, allowing one \textbf{to continuously change embedding vectors in order to decode them and sample new molecular structures.}
Structural similarity (and, therefore, functional similarity) here is maintained by small changes around reference embeddings of known inhibitors.
New molecules can be exposed to a number of rules evaluating the quality of samples.
In drug design, these rules may include drug-likeness, toxicity, inhibitory action towards a target protein, synthetic accessibility, etc.
These rules are based on different quantities, which in part are fully deterministic (like molecular weight or number of heavy atoms), some quantities are empirically parameterized (e.g., lipophilicity), some (for example, inhibition strength) need to be predicted via machine learning approaches.

We already mentioned the low availability of inhibitory data about coagulants, thus, we will apply the scheme for coagulants and 
anticoagulants -- the latter ones will help validate the scheme.
Generally, the inhibitory data often appear in two types (see the \hyperref[sec:Data]{Data} section): binary class (active \textit{vs} inactive) and continuous values reflecting inhibitory strength  (inhibition constants $K_i$, half-inhibitory concentrations IC50, dissociation constants $K_d$, etc.) of a molecule towards a chosen protein.
In order to use all the available data for anticoagulants, we apply a two-stage assessment: (i) by predicting an inhibitory class and (ii) if the class is active, by predicting inhibition constant $K_i$.
In our setup, both stages consist of multiple ensembled machine learning models to make predictions more robust.~\cite{Druchok2021}
Predicted $K_i$ values allow us to define a ``radius'' around reference inhibitors, where the generated structures demonstrate an acceptable inhibition action.
The autoencoder architecture, assessment rules, and sampling strategies are discussed in the following subsections.

\subsection*{\sffamily \large Autoencoder}

Principal dimensionalities of autoencoder are defined by the nature of data for known Thrombin and Protein C inhibitors.
In our approach, SMILES notations are one-hot encoded, producing 2D matrices with sizes defined by the maximal length of SMILES and the size of the dictionary of unique elements.
Before setting the maximal length of SMILES, we would like to tackle the aspect of aromaticity in SMILES -- any molecule, that can be written using aromatic flags, can also be written in Kekul\'e form.
The Kekul\'e form expresses the electronic structure of a molecule using bond types having integer formal bond order, while the aromatic form permits bond elision by introducing additional lowercase atom notations.
Hence, there is a trade-off between a more compact aromatic form with a larger dictionary \textit{vs} slightly longer Kekul\'e notation but a smaller dictionary.
We experimented with both forms and found that the task of training autoencoder is simpler when the Kekul\'e convention is adopted.

The procedure of data acquisition is further discussed in \hyperref[sec:Data]{Data}, while at this stage we need to note that the vast majority of known inhibitors of interest have kekulized SMILES of lengths below 150 elements.
We also limited the dictionary to the following 21 symbols: 
`0', `1', `2', `3', `4', `5', `6',
`=', `\#', `(', `)',
`C', `N', `O', `P', `S',
`Na', `F', `Cl', `Br', `I'.
Hence, the autoencoder in our approach operates with one-hot encodings of the size of $150\times21$.
SMILES samples longer than 150 elements are not considered, while shorter ones are zero-padded.
Zero-padding helps unify the size of input data, but, in the case of constant one-sided zero-padding -- for example, zeros are added to the end of SMILES string only -- it poses a problem of excessive learning of zero values for a part of output neurons.
As a result, longer SMILES sequences tend to be weakly reconstructed when passing through the autoencoder.
In order to overcome this problem we (i) pre-populated the training dataset with SMILES of lengths above half of the maximal length and (ii) introduced left, right, and left-and-right zero-paddings.
The approach with different zero-paddings we successfully applied in Ref.~\citenum{Druchok2022}.

\begin{figure}
\begin{center}
\includegraphics[clip=true,width=0.8\textwidth]{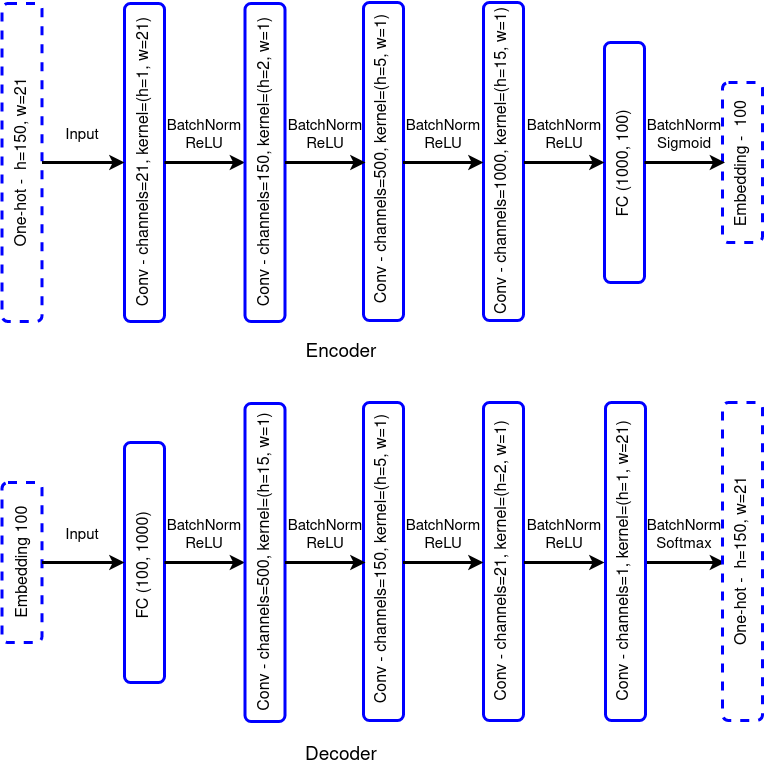}
\caption{The architecture of the Autoencoder. The upper chart presents the Encoder: connections between layers are indicated by black arrows with BatchNorm and activation functions denoted. Each convolution layer is characterized by the number of output channels and kernel sizes (height -- h, width -- w). Solid blue frames indicate the neural layers and the dashed blue frames -- input and output vectors. The bottom chart shows the Decoder scheme with the same notation and color conventions.}
\label{fig:autoencoder}
\end{center}
\end{figure}

The autoencoder architecture was tuned by changing the number of layers, their types, shapes, and activation functions.
We also tested the application of dropouts, but the optimal performance was achieved without it.
The encoder part consists of four convolutional and one fully connected layers.
The decoder part has a mirrored to the encoder structure with one fully connected and four convolutional layers.
All layers were followed by batch-normalization and ReLU activation functions, except for sigmoid and softmax for the last encoder and decoder layers, correspondingly.
The output of the encoder is a vector of 100 float numbers varying between 0 and 1; it is considered to be the embedding vector of chemical compounds.
A layer-by-layer scheme for encoder and decoder is shown in Fig.~\ref{fig:autoencoder}.
The code for the autoencoder is available at \url{https://github.com/mksmd/AE100\_for\_coagulants}.

\subsection*{\sffamily \large Clustering in the space of embeddings}

We already mentioned that our aim is to generate candidate molecules with properties close to reference ones. 
Following the assumption that structurally similar molecules demonstrate similar properties, we analyze the configuration of known coagulants and anticoagulants in the space of embeddings based on the Euclidean distance. 
Given two embeddings $a$ and $b$, the Euclidean distance is defined by:
\begin{align}
    d_{a-b} = 
    \left[
    \sum_{i=1}^n \left(x_i^{(a)} - x_i^{(b)}\right)^2
    \right]^{1/2},
\end{align}
where $n=100$ is the space dimensionality and $x_i^{(a, b)}$ stand for $i$-th coordinate of a corresponding embedding vector.

We expect the Thrombin and Protein C inhibitors to group in clusters rather than to constitute a solid domain. 
Comparing mutual distances for inhibitors within clusters and distances between Protein C inhibitors and Thrombin inhibitors, we are able to define a characteristic scale $d_{\rm sep}$ corresponding to the best separability of the two inhibitor types in the embedding space.

\subsection*{\sffamily \large Generation of candidates}

\begin{figure}[h!]
    \centering
    \includegraphics[scale=0.75]{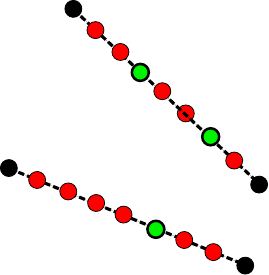}
\qquad\qquad
    \includegraphics[scale=0.75]{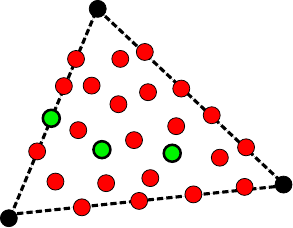}
\qquad\qquad
    \includegraphics[scale=0.75]{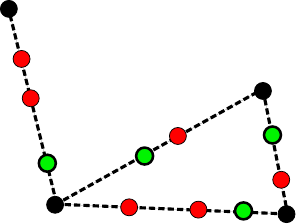}
    \caption{Generation of new embeddings using different interpolation techniques: the closest pairs (left panel), a triangle (middle panel), and SMOTE (right panel). Reference embeddings (of known inhibitors) are denoted by filled black circles. For pairs and SMOTE, dotted lines indicate the exploration paths. For the triangle, dotted lines delimit the interpolation domain (edges are included). Embeddings with incorrect SMILES (red) are filtered out; only the correct ones (green) are retained.}
    \label{fig:smote_emb}
\end{figure}

\textbf{Interpolation between known inhibitors}.
Depending on the distribution of inhibitors in the embedding space, we can consider several strategies to generate new compounds by interpolation between reference points.
The simplest approach would be to homogeneously probe edges between two known inhibitors with a small step.
Another option would be probing multiple points in two-dimensional domains defined by three known ligands.
Alternatively, one can engage Synthetic Minority Over-sampling Technique (SMOTE), a method used to address the problem of imbalanced datasets in machine learning.~\cite{Chawla2002,imblearn:www} 
In our case, such a technique is applied to generate additional data similar to known entities based on a limited set of the latter.
The interpolation approaches are sketched in Fig.~\ref{fig:smote_emb}.  

\begin{figure}
    \centering
    \includegraphics[scale=0.75]{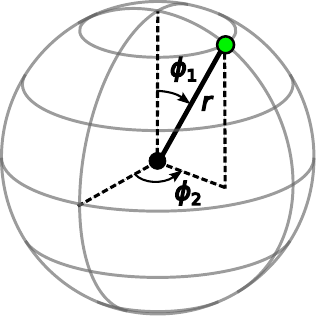}
\qquad
    \includegraphics[scale=0.75]{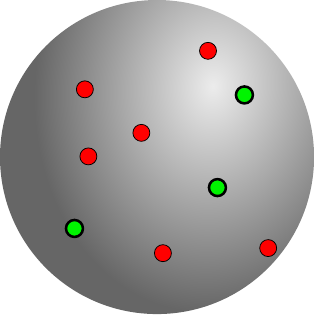}
    \caption{Left panel: Embedding of the reference inhibitor (full black circle) defines center of a hypersphere, while point on the sphere (full green) surface with the radial ($r$) and angular ($\phi_1, \phi_2$) coordinates corresponds to generated embedding.
    Right panel: Embeddings with incorrect SMILES (red) are filtered out and only the correct ones (green) are retained.}
    \label{fig:sphere_emb}
\end{figure}

\textbf{Hypersphere search}. New candidate molecules can be sought for in the embedding space around the known inhibitors. 
It seems reasonable to assume that the highest likelihood of locating them corresponds to distances $r < d_{\rm sep}$. 
In order to perform such a search, it is convenient to use the hyperspherical coordinates, as demonstrated in Fig.~\ref{fig:sphere_emb} on a simplified example of an ordinary three-dimensional sphere.
For every radius $r_0$, $r_1 = r_0 + \Delta r$,\ldots, $r_k = r_0 + k\Delta r$, \ldots\ up to $d_{\rm sep}$ we randomly generate $N$ values of hyperangles $\phi_1,\ldots, \phi_{n-1}$.
The first $(n-2)$ angles $\phi_i$ vary within $[0, \pi)$ and the last one $\phi_{n-1}$ lies within $[0, 2\pi)$.
Conversion from hyperspherical to Cartesian coordinates is then done using the following formulas:
\begin{align*}
x_1   &= r \cos \phi_1,\\
x_2   &= r \sin \phi_1 \cos \phi_2,\\
x_3   &= r \sin \phi_1 \sin \phi_2 \cos \phi_3,\\
\ldots\\
x_{n-1} &= r \sin \phi_1 ... \sin \phi_{n-2} \cos \phi_{n-1},\\
x_n   &= r \sin \phi_1 ... \sin \phi_{n-2} \sin \phi_{n-1}.
\end{align*}
With the known inhibitor at the center of the hypersphere having the embedding vector $(X_1^{\rm orig}, \ldots, X_n^{\rm orig})$, the coordinates in generated vectors are thus given by
$$
(X_1^{\rm orig}+x_1, \ \ldots, \ X_n^{\rm orig}+x_n).
$$

The generated samples can be further filtered using additional constraints based on the proximity of the desired/undesired inhibitor groups in the embedding space.
Finally, the generated embedding vectors are fed to the decoder to obtain SMILES and examine them for validity.
Note that due to the continuous nature of the embedding space, multiple embeddings can correspond to one SMILES.

\subsection*{\sffamily \large Filters for the generated molecules}

The list of drug candidates can be exposed to additional filters to rank and prioritize them.
A possible approach is to estimate drug-likeliness, and to this end, a number of filters based on pharmacokinetic principles can be applied.~\cite{Pathania2021,Kralj2022}
Such filters are often based on molecular weight, $n$-octanol/water partition coefficient $\log P$, degree of polarity, and number of bonds, atoms, hydrogen bond donors/acceptors, etc.
We also analyze the Synthetic Accessibility Score~\cite{Ertl2009,Daina2017} -- a measure of difficulty to synthesize a particular compound, ranging from 1 to 10, where 1 corresponds to very easy and 10 means very difficult.
These filters, however, do not serve as a decisive prescription and have rather an auxiliary advisory role.
Neither a ligand fulfilling the rules can be automatically considered a proper drug component nor the failure to conform to the rules eliminates this ligand from further evaluation --
for instance, a large share of FDA-approved drugs do not pass the abovementioned filters.~\cite{OHagan2015,Kralj2022}

\subsection*{\sffamily \large Thrombin affinity predictors}

As described in the general outline of \hyperref[sec:Methods]{Methods} section, we apply a two-stage procedure: a candidate is first sieved through a binary active--nonactive classifier, and only those classified active are further sent to the regression stage.
We consider four algorithms for both classification and regression tasks:
Support Vector Machine~(SVM),\cite{Cortes1995} Random Forest~(RF),\cite{Breiman2001} LightGBM~(LGBM),\cite{Ke2017} and XGBoost~(XGB).\cite{Chen2016}

It is a common practice to use molecular fingerprints as numerical representations of small organic molecules.
We have utilized fingerprints alongside the autoencoder embeddings in our analysis.
For each ligand, a SMILES string was obtained in the canonical form, with the stereochemistry information removed. 
The SMILES string was subsequently fed to the RDKFingerprint algorithm from the RDKit library~\cite{RDKit:www} to generate a fixed-length fingerprint as a binary string.
The fingerprint length was set to the default value of 2048 bits.
The obtained strings are the so-called daylight-like topological fingerprints.~\cite{Green2021}
Despite a widespread use of fingerprints, there is no efficient way to reconstruct SMILES back from them.
As one can use either embeddings or fingerprints as inputs to predictive models, it is worth unveiling some results that influenced the choice of methods in our study.
In particular, with the SVM classification model, we were able to achieve a precision score of 0.817 based on embeddings and 0.897 based on fingerprints.
In the SVM regression model, the $R^2$ score of 0.759 was achieved for fingerprints, while a rather low $R^2 = 0.448$ was obtained for embeddings.
The values of the precision and $R^2$ scores were coherently lower for embeddings in comparison to fingerprints in all the analyzed models. 
We, therefore, focus on the fingerprint cases in the description of the classification and regression methods below.

\textbf{Support Vector Machine}.
The RBF kernel was found to yield the best scores in both the classification and regression tasks.
In the classification task based on fingerprints, the grid search for the regularization parameter \texttt{C} was run among the set of values 0.1, 0.5, 1--10 with steps 1, 20, 50, and 100 -- the highest precision score was achieved with \texttt{C} = 2.
For the regression task, the same set of values revealed that \texttt{C} = 3 yields the highest $R^2$ score.

\textbf{Random Forest}.
For the classification model based on fingerprints, the following set of hyperparameters was found optimal:
\texttt{n\_estimators} = 500,
\texttt{max\_depth} = 6,
\texttt{min\_samples\_split} = 2,
and \texttt{ccp\_alpha} = 0.
The set of hyperparameters:
\texttt{n\_estimators} = 200,
\texttt{max\_depth} = 10,
\texttt{min\_samples\_leaf} = 1,
\texttt{min\_samples\_split} = 4,
and \texttt{ccp\_alpha} = 0
was found optimal for the regression model.

\textbf{LightGBM}.
For the classification task, the following set of hyperparameters was found optimal:
\texttt{n\_estimators} = 50,
\texttt{max\_depth} = 4,
\texttt{learning\_rate} = 0.2,
and \texttt{reg\_alpha} = 0.
The optimal hyperparameters for the regression model:
\texttt{n\_estimators} = 200,
\texttt{max\_depth} = 4,
\texttt{num\_leaves} = 20,
\texttt{learning\_rate} = 0.1,
and \texttt{reg\_alpha} = 0.

\textbf{XGBoost}.
For the classification model, the optimal set of hyperparameters appeared to be:
\texttt{n\_estimators} = 200,
\texttt{max\_depth} = 4,
and \texttt{learning\_rate} = 0.05.
The \texttt{objective} parameter was set to `binary:logitraw' ensuring better performance in comparison with the default `binary:logistic' value.
For the regression, the optimal hyperparameters are:
\texttt{n\_estimators} = 200,
\texttt{max\_depth} = 4,
\texttt{learning\_rate} = 0.1,
and \texttt{reg\_alpha} = 1.

\textbf{Combinations of models.} Having the four above-mentioned models, we employed voting scenarios based on various model combinations (two, three, and all four models).
For the classification task, a ligand is assumed to be active if all the involved models provide such a prediction unanimously. 
In the regression task, the arithmetic mean of the predicted value is used.

\subsection*{\sffamily \large MegaMolBART}

We also compare the performance of our autoencoder-based approach with  MegaMolBART, a generative AI model developed jointly by AstraZeneca and NVIDIA.~\cite{Philippidis2021} 
The name originates from Bidirectional and Auto-Regressive Transformer (BART)~\cite{Lewis2020} -- another autoencoder architecture, initially designed to reconstruct textual content.
The MegaMolBART relies on NVIDIA's NeMo Megatron framework and was trained using the ZINC-15 database.~\cite{Sterling2015}
The MegaMolBART (0.1.2 ver.) we used is a part of the Cheminformatics container~\cite{NVIDIA_Cheminformatics} within the NVIDIA Clara Discovery collection.
It allows for the generation of molecules either by interpolating between two reference molecules or by sampling around one reference molecule specified by ChEMBL IDs (version 27 of the ChEMBL database~\cite{Mendez2018,ChEMBL27}).
Apparently, these generation strategies align with the strategies employed within our autoencoder approach.

\section*{\sffamily \Large DATA}\label{sec:Data}

\subsection*{\sffamily \large Data for the autoencoder}

As this research is an extension of our previous study,~\cite{Druchok2021b} to train autoencoder on valid SMILES examples, we used previously downloaded eMolecules dataset~\cite{eMolecules} -- a large set of commercially distributed chemicals listed in the SMILES format.

\subsection*{\sffamily \large Data for activity / inactivity}

The inhibitors dataset we used in the present work comes from the BindingDB database.~\cite{BindingDB:www,Chen2001,Chen2002b,Liu2007,Gilson2015}
It contains information about protein-ligand interactions that can be used to identify potential drug candidates.
We focus on two proteins relevant to blood clotting control, Thrombin and Protein C.
The database contains a significantly larger amount of records for the former, while the number of coagulants (Protein C inhibitor) is rather low. 
Apart from Protein C, there are only 21 ligands for human Antithrombin, of those only six are active based on the values of IC50 (1 entry) and $K_d$ (5 entries).
The database contains no entries for two other clotting regulation targets -- vitamin K-dependent Protein~S and Protein~Z.

To distinguish active and inactive ligands, we used the values of the inhibition constant ($K_i$), dissociation constant ($K_d$), inhibitory concentration 50\%~(IC50), and effective concentration 50\%~(EC50). 
Usually, a common threshold of 10\,000~nM for all four indicators is applied.~\cite{Lee2017,Kowalewski2020,Shim2021,Druchok2021}
However, stronger~\cite{Lee2019} or diversified thresholds, e.\,g., 10\,000~nM for $K_i$ and $K_d$ and 20\,000~nM for IC50 and EC50, can be adopted as well.~\cite{Chan2020} 
We decided to stay within a common threshold of 10\,000~nM and work in terms of $\log_{10}K_i$.
There are several reasons to use $\log_{10}K_i$.
It converts concentrations differing by several orders of magnitude to a more homogeneous domain, which makes machine learning training on different concentration scales more efficient.
By taking a logarithm, we also implicitly account for the relevance of relative errors in experimental measurements of $K_i$ rather than absolute ones.
So, the upper threshold for active inhibitors in the present work is $\log_{10}K_i \leq 4$.
Ligands with values $\log_{10}K_i > 6$ were excluded from the training due to their limited clinical usefulness.
The BindingDB records for Thrombin and Protein C were cleared from duplicates and SMILES that cannot be canonicalized correctly.
Then SMILES with length above 150 and those containing elements beyond the dictionary were removed.
Finally, all the records lacking at least one of the binding parameters ($K_i$, $K_d$, EC50, or IC50) were filtered out.

\begin{figure}
\centering
\includegraphics[clip=true,scale=0.55]{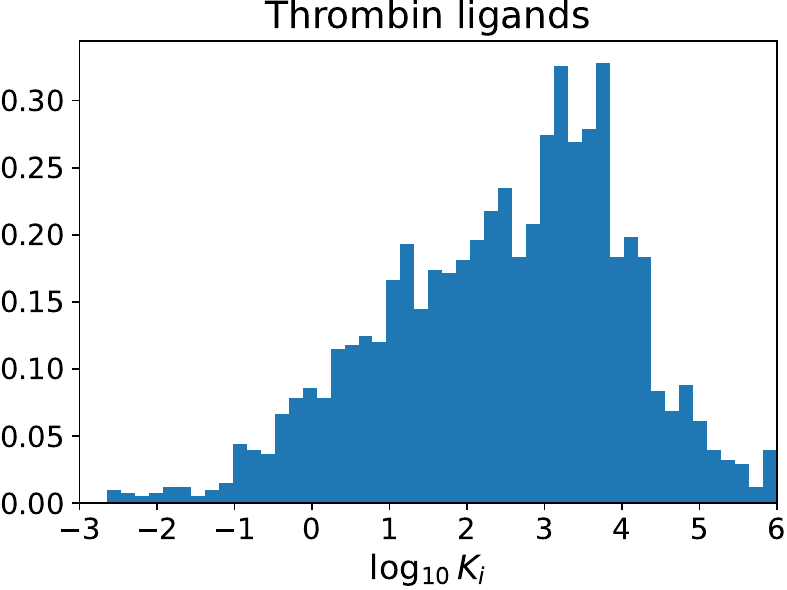}
\includegraphics[clip=true,scale=0.55]{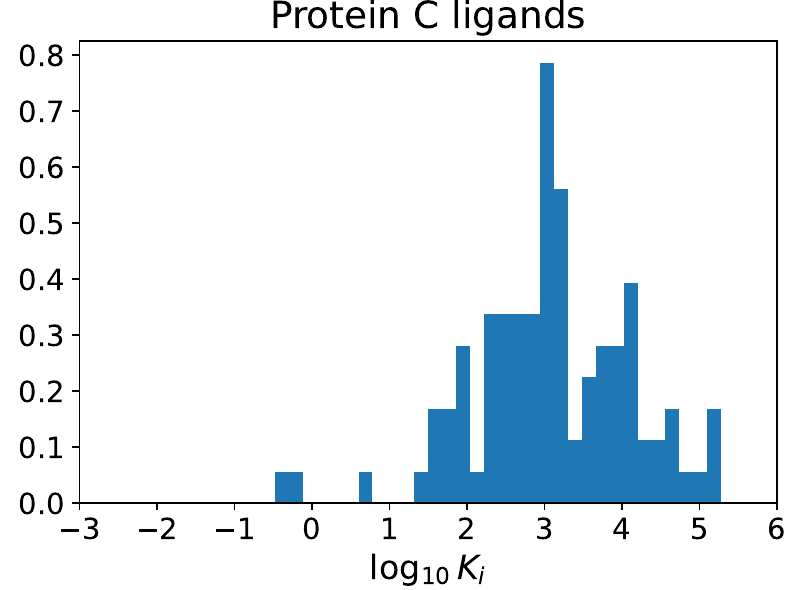}
\caption{Distribution of the $\log_{10} K_i$ values within the Thrombin (left) and Protein C (right) datasets used in the regression training.}
\label{fig:logKi_hist_thrombin_protc}
\end{figure}

\textbf{Thrombin.}
The classification dataset contains 5009 ligands, of those 3323 are active as based on the threshold of 10\,000~nM for $K_i$, $K_d$, EC50, or IC50. 
The regression dataset consists of 2270 ligands with available $K_i$ values, of those 1927 active (with $K_i \leq 10\,000$~nM).
The $\log_{10} K_i$ distribution for Thrombin ligands is shown in the left panel of Fig.~\ref{fig:logKi_hist_thrombin_protc}.

\textbf{Protein C.}
BindingDB contains 188 Protein C ligands. 
Of those, 103 ligands are classified as active and 85 as inactive, respectively, while $K_i$ is available for 125 ligands only.
For 26 out of 125, $K_i$ is specified in the form of lower bound, e.\,g., ``$>\!1000$'', ``$>\!18000$'', or ``$>\!37600$'', hence, they were filtered out to avoid uncertainties.
Of the remaining 99 ligands with $K_i$ specified, 79 are active Protein C inhibitors (with $K_i \leq 10\,000$\,nM).
The corresponding $\log_{10} K_i$ distribution for Protein C ligands is shown in the right panel of Fig.~\ref{fig:logKi_hist_thrombin_protc}. 

\section*{\sffamily \Large RESULTS}\label{sec:Results}

\subsection*{\sffamily \large Autoencoder training}

\begin{figure}
\begin{center}
\includegraphics[clip=true,scale=0.55]{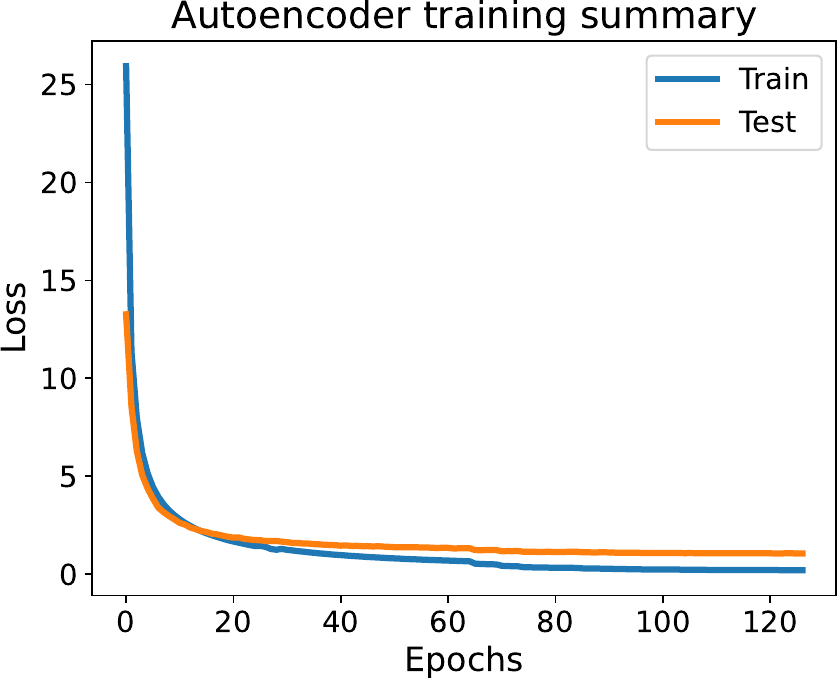}
\caption{The loss function on train (blue) and test (orange) subsets along the autoencoder training process.}
\label{fig:ae_training}
\end{center}
\end{figure}

Training of the autoencoder took 127 epochs before it got terminated by an early stopping condition.
This condition is satisfied if no loss improvement on the test subset was achieved during six consecutive epochs.
The initial value for the learning rate was set to 0.001 and decreased by a factor of 0.7 after each three consecutive epochs with no loss improvement on the test subset.
The learning rate before termination had reached the value of $\approx0.00002$.
In Fig.~\ref{fig:ae_training} we show the evolution of the loss function during the training process.
The model with the best test loss was then used in all further experiments.

\subsection*{\sffamily \large Training of Thrombin affinty predictors}

In order to select the most relevant metric for the classification problem, let us briefly remind the reader of the definitions of some popular scores:
\begin{align*}
&{\rm 
Precision = \frac{True\ Positive}{Total\ Predicted\ Positive}
 = \frac{TP}{TP \, + \, FP}
},\\[12pt] 
&{\rm 
Accuracy = \frac{Correct\ Predictions}{All\ Predictions}
 = \frac{TP \, + \, TN}{TP \, + \, TN \, +\, FP \, +\, FN}
},\\[12pt]
&{\rm 
Recall = \frac{True\ Positive}{Total\ Actual\ Positive}
 = \frac{TP}{TP \, + \, FN}
},
\end{align*}
where the standard abbreviations are used: TP for true positive, TN for true negative, FP for false positive, and FN for false negative, respectively.
As we aim at a correct identification of active inhibitors (positive instances) minimizing the number of incorrect positive predictions, the precision score is best suited for such a task.
We, thus, focus on this score in the classification.

The resulting scores for all the four models are shown in Table~\ref{tab:class}. 
As we have mentioned in the \hyperref[sec:Methods]{Methods} section, the obtained scores based on embeddings are systematically lower than those for fingerprints, thus, further, we will present the classification/regression results based on fingerprints only.
Among SVM, RF, LGBM, and XGB, the best precision is demonstrated by the XGB model.
For the classification task, we have also applied combinations of models. 
In such cases, an item is classified as active only if all the models unanimously predict so. 
Obviously, the highest precision is achieved when all the four models are combined, see Table~\ref{tab:class}. 
\begin{table}
\caption{Precision scores for classification with models and their combinations}\label{tab:class}
\medskip
\centering
\begin{tabular}{lc}
\hline\noalign{\smallskip}
Model combination & Precision \\
\noalign{\smallskip}\hline\noalign{\smallskip}
SVM             & 0.897 \\
RF              & 0.893 \\
LGBM            & 0.877 \\
XGB             & 0.912 \\
SVM + RF        & 0.919\\
SVM + LGBM      & 0.912\\
SVM + XGB       & 0.921\\
RF + LGBM       & 0.912\\
RF + XGB        & 0.923\\
LGBM + XGB      & 0.917\\
SVM + RF + LGBM & 0.925\\
SVM + RF + XGB  & 0.929\\
SVM + LGBM + XGB & 0.924\\
RF + LGBM + XGB & 0.926\\
\cellcolor{lightgray}\textbf{SVM + RF + LGBM + XGB} & \cellcolor{lightgray}\textbf{0.930}\\
\noalign{\smallskip}\hline
\end{tabular}
\end{table}
For fingerprints, the precision score is 0.930 while for embeddings its highest value is 0.843 also with four models applied simultaneously.

\begin{figure}
    \centering
    \includegraphics[scale=0.55]{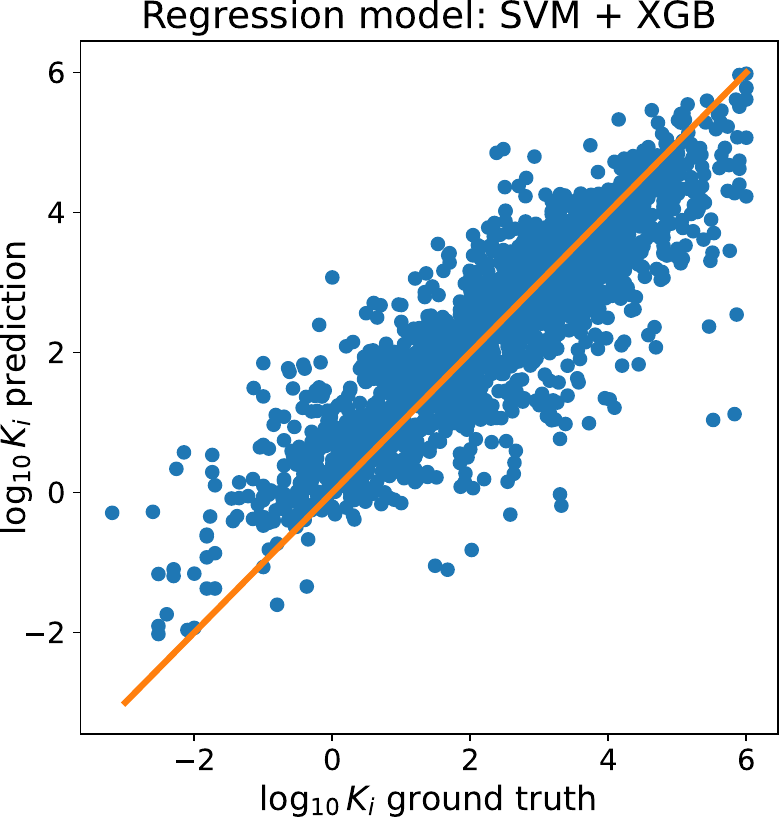}
    \caption{Comparison of actual versus predicted binding affinities for known Thrombin ligands by the best-performing model combination (SVM + XGBoost). The orange line corresponding to ideal reconstruction is intended to guide the eye.}
    \label{fig:regresion_thrombin_best}
\end{figure}

Results of the regression analysis for fingerprints are given in Table~\ref{tab:regr}. 
\begin{table}
\caption{Regression scores for models and their combinations}\label{tab:regr}

\medskip
\centering
\begin{tabular}{lccc}
\hline
Model combination & $R^2$ & MSE & MAE \\
\hline
SVM             & 0.759 & 0.590 & 0.545 \\
RF              & 0.734 & 0.652 & 0.586 \\
LGBM            & 0.750 & 0.612 & 0.567 \\
XGB             & 0.754 & 0.601 & 0.562 \\
SVM + RF          & 0.755 & 0.600 & 0.556 \\
SVM + LGBM        & 0.762 & 0.583 & 0.547 \\
\cellcolor{lightgray}\textbf{SVM + XGB} & \cellcolor{lightgray}\textbf{0.764} & \cellcolor{lightgray}\textbf{0.578} & \cellcolor{lightgray}\textbf{0.544} \\
RF + LGBM         & 0.752 & 0.608 & 0.564 \\
RF + XGB          & 0.753 & 0.604 & 0.562 \\
LGBM + XGB        & 0.756 & 0.598 & 0.560 \\
SVM + RF + LGBM     & 0.759 & 0.590 & 0.552 \\
SVM + RF + XGB      & 0.760 & 0.587 & 0.551 \\
SVM + LGBM + XGB    & 0.763 & 0.581 & 0.548 \\
RF + LGBM + XGB     & 0.756 & 0.597 & 0.559 \\
SVM + RF + LGBM + XGB & 0.761 & 0.586 & 0.551 \\
\hline
\end{tabular}
\end{table}
The SVM, LGBM, and XGB models demonstrate close values of the $R^2$ score, while the RF model underperforms slightly.
The combination of SVM and XGB models yields both the highest $R^2$ score and the least MSE and MAE; cf. also Fig.~\ref{fig:regresion_thrombin_best}.
For embeddings, the highest $R^2 = 0.448$ was achieved for the SVM model with the remaining three not reaching 0.4.

Thus, for the subsequent predictions, we will rely on best-performing model combinations (highlighted with grey in Tables~\ref{tab:class} and \ref{tab:regr}) based on fingerprints.

\subsection*{\sffamily \large Clustering for Thrombin and Protein C}

\begin{figure}[h]
    \centering
    \includegraphics[clip=true,scale=0.55]{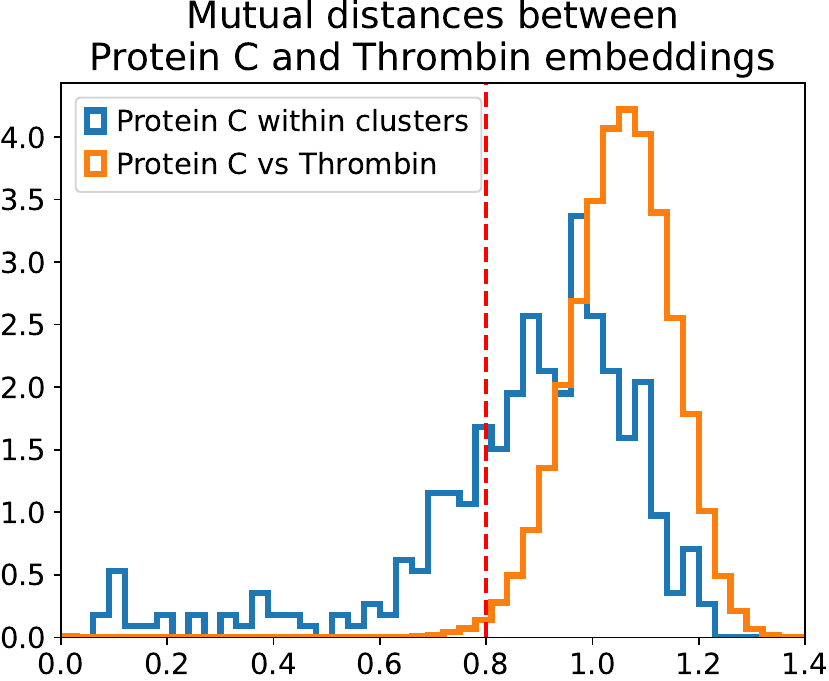} \ 
    \includegraphics[clip=true,scale=0.55]{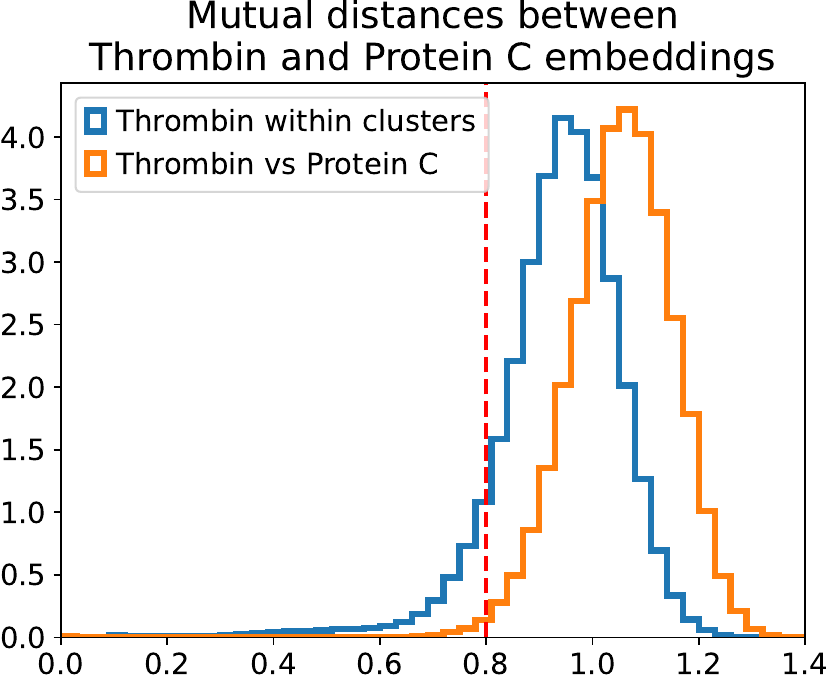}
    \caption{Mutual distances between Protein C and Thrombin inhibitors}
    \label{fig:mutual_dist}
\end{figure}

The cluster analysis based on the Euclidean distances between the embedding vectors revealed ten clusters for active Protein C inhibitors and nine clusters for active Thrombin inhibitors. 
We compared the distances within clusters of Protein C with distances between all Protein C and all Thrombin inhibitors. 
A similar procedure was done in the opposite direction, for distances within the Thrombin clusters versus distances between all active Thrombin and Protein C inhibitors. The results are shown in Fig.~\ref{fig:mutual_dist}.
Based on Fig.~\ref{fig:mutual_dist}, we chose $d_{\rm sep} = 0.8$.
To remind the reader, $d_{\rm sep}$ is intended to serve as a ``safe'' radius around known inhibitors -- other inhibitors of the same protein are spotted there, while the opposite protein inhibitors are weakly populated.

\subsection*{\sffamily \large Generation of candidates for Thrombin inhibitors}

\textbf{Interpolation.}
The SMOTE approach was used to generate 200\,000
embeddings with the \texttt{k\_neighbors} parameter equal to 20.
Out of 200\,000 embeddings, 32\,771 correct SMILES were obtained corresponding to 5736 unique molecules. 
After removing the original Thrombin inhibitors from the generated set, 5354 unique SMILES remained.
Upon accepting those with a distance to the nearest known Thrombin inhibitor below 0.5 and larger than 0.7 to the nearest strong Protein C inhibitor (with $\log_{10}K_i\leq2$), we ended up with 14\,074 embeddings corresponding to 3150 unique SMILES, with 1420 of them predicted active.
These threshold distances, obtained upon a series of trials, maintain a reasonable balance between the numbers of generated accepted and rejected compounds.

\textbf{Hypersphere search.}
Around each of the 2270 Thrombin ligands, hyperspheres were drawn with radii from a half-open interval [0.1, 0.8) with step 0.02. 
Ten random points were cast for every radius value. 
The total number of the generated embedding vectors was thus 794\,500.
Upon filtering out duplicates, 255\,012 embeddings converting to correct SMILES were identified;
202\,132 of them remained after the original ligands were excluded.
These correspond to 34\,957 unique SMILES.
The acceptance criteria based on distances to the known Thrombin and Protein C inhibitors were satisfied by 90\,467 embeddings corresponding to 16\,027 unique SMILES.  
Of those, 6369 unique SMILES were predicted active. The above numbers are summarized in Table~\ref{tab:thromnin_generated_summary}

The lists of accepted and active molecules from SMOTE and hypersphere have the intersection on 673 SMILES.
Thus, in total, the two lists thus contain 7116 unique SMILES.

\begin{figure}[h]
    \centering
    \includegraphics[clip=true,scale=0.55]{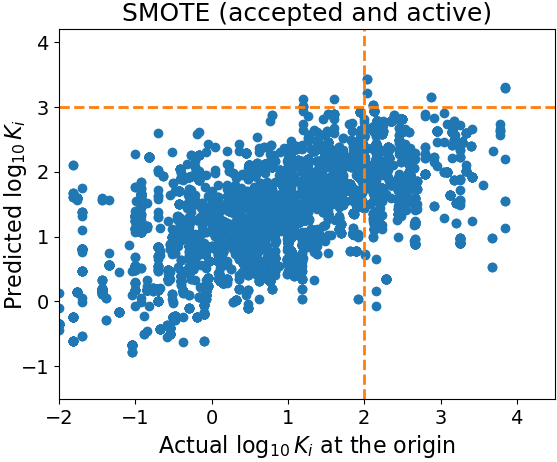} \ 
    \includegraphics[clip=true,scale=0.55]{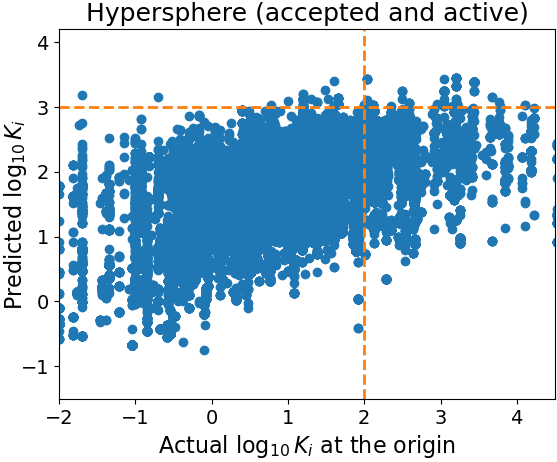}
    \caption{Values of $\log_{10}K_i$ (predicted versus actual at the origin) for Thrombin inhibitors generated using the interpolation by SMOTE and hypersphere approaches}
    \label{fig:gen_kpred_vs_korig}
\end{figure}

To justify the choice of reference points for generation, we analyze the distribution of predicted $\log_{10}K_i$ values for generated compounds as a function of actual $\log_{10}K_i$ for reference inhibitors.
First, generated compounds were assessed for their activity class with the ensemble of classification models SVM~+~RF~+~LGBM~+~XGB, then, the compounds predicted active were fed to the ensemble of regression models SVM~+~XGB predicting their $\log_{10}K_i$ values.
Fig.~\ref{fig:gen_kpred_vs_korig} demonstrates the resulting $\log_{10}K_i$ dependencies for interpolation and hypersphere search around Thrombin inhibitors.
Based on these plots (see dashed line delimiters), in order to ensure that the $\log_{10}K_i$ values for the generated inhibitors stay well below 4, one needs to select reference inhibitors with $\log_{10}K_i \leq 2$.

\begin{table}[h]
    \caption{The summary of generated embeddings and unique SMILES for Thrombin: ``all'' corresponds to correct SMILES, ``accepted'' denotes those satisfying both distance conditions to nearest Thrombin and Protein C inhibitors, and ``active'' as based on the two-stage prediction.}
    \label{tab:thromnin_generated_summary}
    \smallskip
    \centering
\begin{tabular}{l|cc}
\hline
\diagbox[width=2.22in, height=6ex]{Approach}{Types}
 & Embeddings & SMILES \\
\hline
Interpolation (SMOTE); all & 24\,407 & 5354 \\
Interpolation (SMOTE); accepted & 14\,074 & 3150 \\
Interpolation (SMOTE); accepted active & 7282 & 1420 \\
\hline
Hypersphere; all & 202\,132 & 34\,957 \\
Hypersphere; accepted & 90\,467 & 16\,027\\
Hypersphere; accepted active & 48\,683 & 6369 \\
\hline
\end{tabular}
\end{table}

\subsection*{\sffamily \large Generation of candidates for Protein C inhibitors}

The reported results for generated Thrombin inhibitors provide some hints on how potential Protein C inhibitors should be sought.
Symmetrically to the Thrombin generation case, we restrict ourselves to strong Protein C inhibitors, with low $\log_{10} K_i \leq 2$ for both interpolation and hypersphere search. 
Of 79 active Protein C inhibitors, only 15 satisfy this stronger condition; they are listed in Table~\ref{tab:strongProtC-img}.

\def\imgscale{0.21}

\begin{table}[p]
\caption{Strong Protein C inhibitors.}\label{tab:strongProtC-img}

\smallskip
    \centering
\begin{tabular}{cccccc}
\hline
$s$ & $\log_{10} K_i$ & $s$ & $\log_{10} K_i$ & $s$ & $\log_{10} K_i$ \\
\hline
 1 & $-0.34$ &
 2 & $-0.21$ &
 3 & $ 0.71$ \\
\multicolumn{2}{c}{\includegraphics[scale=\imgscale]{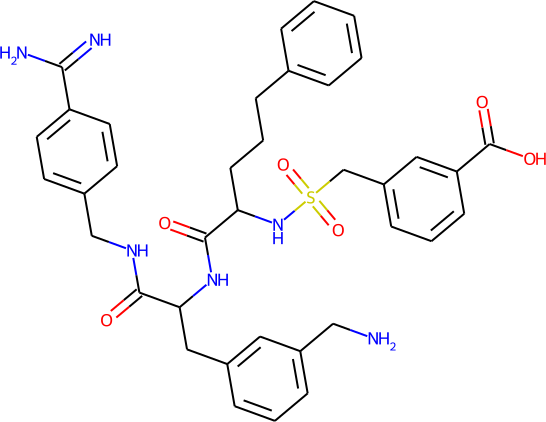}} & \multicolumn{2}{c}{\includegraphics[scale=\imgscale]{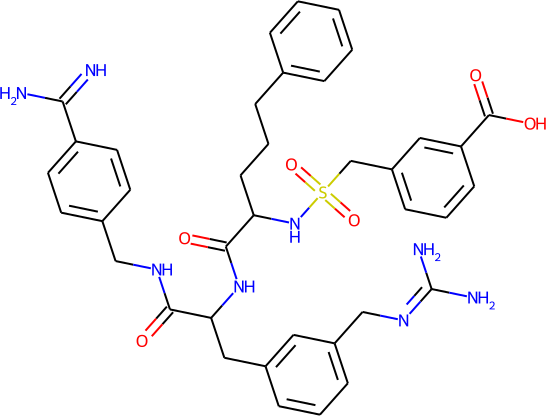}} & \multicolumn{2}{c}{\includegraphics[scale=\imgscale]{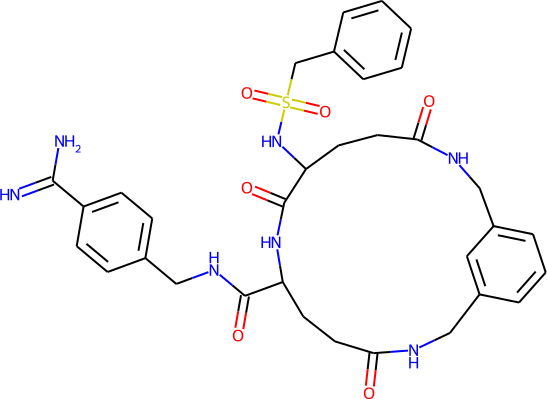}} \\
 4 & $ 1.43$ &
 5 & $ 1.59$ &
 6 & $ 1.60$ \\
\multicolumn{2}{c}{\includegraphics[scale=\imgscale]{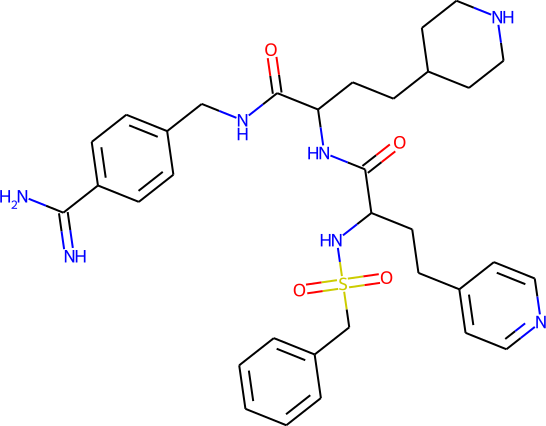}} & \multicolumn{2}{c}{\includegraphics[scale=\imgscale]{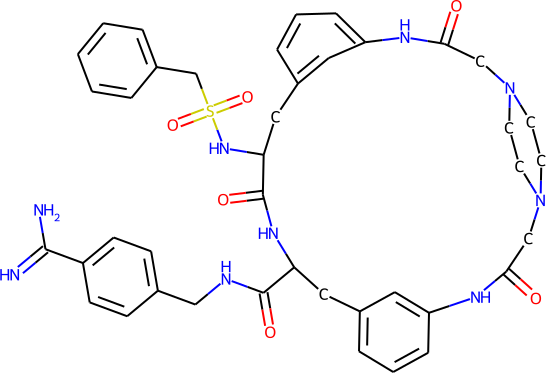}} & \multicolumn{2}{c}{\includegraphics[scale=\imgscale]{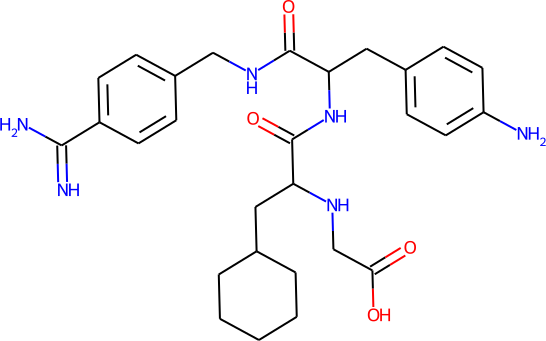}} \\
 7 & $ 1.61$ &
 8 & $ 1.68$ &
 9 & $ 1.78$ \\
\multicolumn{2}{c}{\includegraphics[scale=\imgscale]{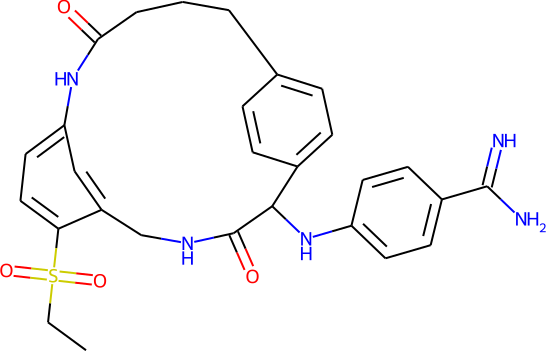}} & \multicolumn{2}{c}{\includegraphics[scale=\imgscale]{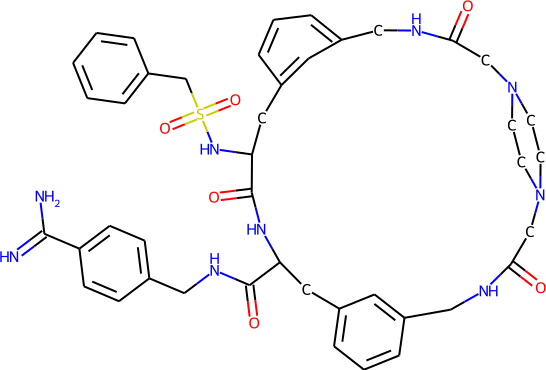}} & \multicolumn{2}{c}{\includegraphics[scale=\imgscale]{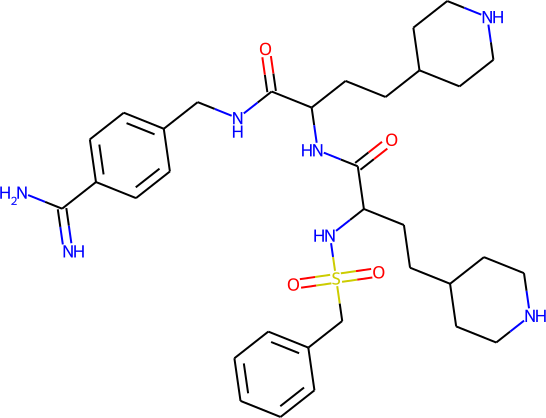}} \\
10 & $ 1.85$ &
11 & $ 1.86$ &
12 & $ 1.94$ \\
\multicolumn{2}{c}{\includegraphics[scale=\imgscale]{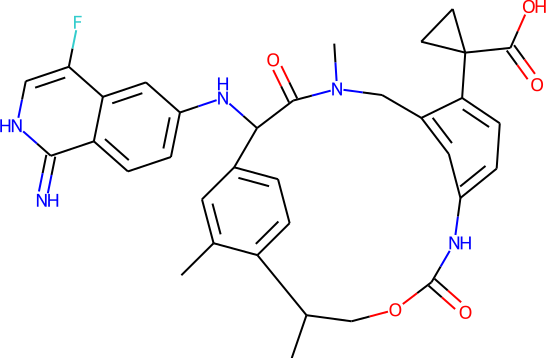}} & \multicolumn{2}{c}{\includegraphics[scale=\imgscale]{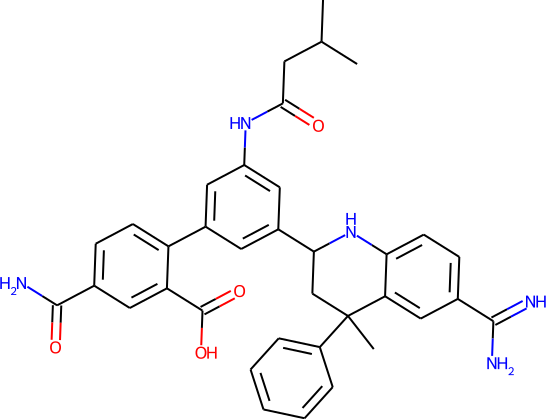}} & \multicolumn{2}{c}{\includegraphics[scale=\imgscale]{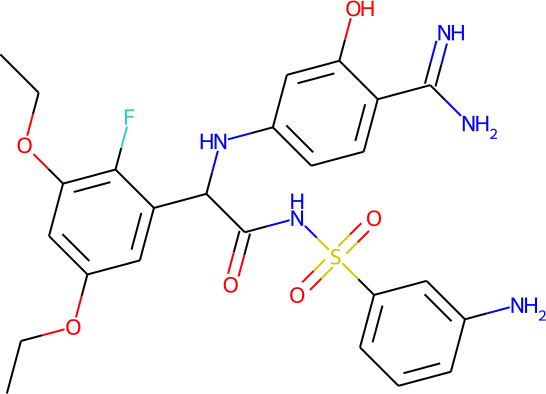}} \\
13 & $ 1.95$ &
14 & $ 2.00$ &
15 & $ 2.00$ \\
\multicolumn{2}{c}{\includegraphics[scale=\imgscale]{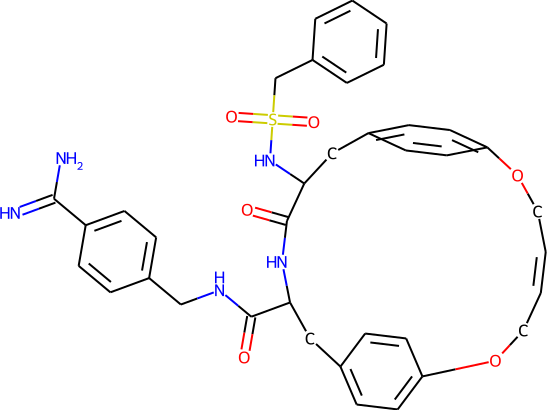}} & \multicolumn{2}{c}{\includegraphics[scale=\imgscale]{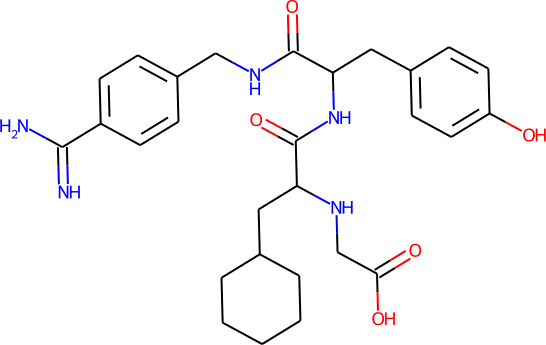}} & \multicolumn{2}{c}{\includegraphics[scale=\imgscale]{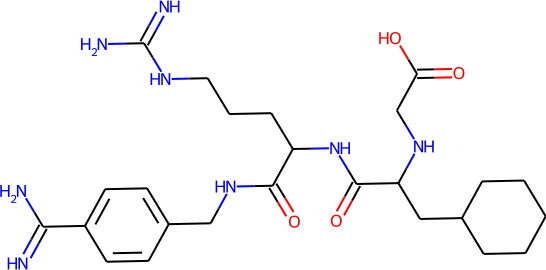}} \\
\hline
\end{tabular}
\end{table}

\textbf{Interpolation with SMOTE.}
Using the SMOTE approach, we generated 100\,000 embeddings interpolating all the 15 strong Protein C inhibitors, i.\,e., with $\texttt{k\_neighbors} = 14$.
Upon applying this procedure, we obtained 6568 embeddings corresponding to 89 correct and unique SMILES.
This set was filtered according to the following acceptance criteria:
distance to the nearest strong Protein C inhibitor should be below 0.5 and larger than 0.7 from the nearest Thrombin inhibitors (with $\log_{10} K_i\leq2$).
Such a filter imposes additional restrictions that aim at the identification of new ligands being more similar to the known Protein C inhibitors and, at the same time, dissimilar from the Thrombin ones.
Only 2337 embeddings satisfy these acceptance criteria yielding 30 unique SMILES from the SMOTE analysis.

\textbf{Interpolation within the strongest triangle.}
As Table~\ref{tab:strongProtC-img} shows, Protein C inhibitors with $s=1, 2,$ and 3 are the strongest ones.
In the embedding space, they form an almost equilateral triangle with mutual distances as follows:
\begin{itemize}
    \item $s=1$ to $s=2$: $d = 0.953$,
    \item $s=1$ to $s=3$: $d = 1.069$.
    \item $s=2$ to $s=3$: $d = 1.005$,
\end{itemize}
The circumradius of this triangle is 0.585, thus, potentially, any new instances within it can be close to at least two or even three strongest Protein C inhibitors.
Upon casting one million random points uniformly~\cite{Osada2002} within the triangle, we obtained 6781 embeddings corresponding to correct SMILES, with 13 unique ones.
We have also analyzed in detail the triangle edges with a step of 0.0001 and discovered that only one edge (between $s=2$ and $s=3$) contains embeddings corresponding to correct SMILES.
All the three SMILES from this edge had been already generated within the triangle.
After filtering according to the acceptance criteria, 5620 embeddings remained yielding 6 unique SMILES.

\textbf{Interpolation between closest pairs.}
Mutual distances in the embedding space between the following five pairs of the strong Protein C inhibitors are below the $d_{\rm sep} = 0.8$ threshold:
(3, 4), (4, 9), (6, 13), (6, 14), and (13, 14).
The respective edges were probed with the search step of 0.0001.
This procedure yielded 16508 embeddings corresponding to 14 correct unique SMILES.
Two of these SMILES coincide with those obtained from the strongest triangle as described above.
After filtering according to the acceptance criteria of distance to Thrombin inhibitors, 7569 embeddings remained, yielding 10 unique SMILES.

\textbf{Hypersphere search.}
Around each of 15 strong Protein C inhibitors, hyperspheres were drawn with radii from a half-open interval [0.1, 0.8) with step 0.01, and 100 random points were cast for every radius value. 
The total number of the generated embedding vectors was thus 105\,000. 
The amount of different embedding vectors that yielded correct SMILES was 19\,603.
Out of these, the number of unique SMILES was 2068; none of them coincided with the reference strong Protein C inhibitors.
After filtering according to the acceptance criteria, 15\,126 embeddings remained corresponding to 1545 unique SMILES.

A generation summary for Protein C is shown in Table~\ref{tab:summary_generated}.
A few compounds were generated simultaneously by different approaches.
The total number of unique generated SMILES strings for the Protein C inhibitors is 1571.
Of those, 1545 are unique to the hypersphere set, 18 -- unique to the SMOTE set, 5 -- unique to the closest pairs set, and 3 are unique to the strongest triangle. 
One molecule appeared to be common to all four sets.

Note that, unlike Table~\ref{tab:thromnin_generated_summary},no information about the number of active compounds is provided in Table~\ref{tab:summary_generated}.
In comparison to Thrombin, the binding affinity data for Protein C are too scarce to train an ML classifier, not even mentioning a regressor, and this very fact triggered the presented study.
However, replicating the correlation observed in Thrombin, we expect the compounds generated here to be active Protein C ligands since the reference points are strong Protein C inhibitors with $\log_{10}K_i\leq2$.

\begin{table}[h]
    \caption{The summary of generated embeddings and unique SMILES for Protein~C: ``all'' and ``accepted'' have the same meaning as in Table~\ref{tab:thromnin_generated_summary}.}
    \label{tab:summary_generated}
    \smallskip
    \centering

\begin{tabular}{l|cc}
\hline
\diagbox[width=2.7in, height=6ex]{Approach}{Types} & Embeddings & SMILES \\
\hline
Interpolation (SMOTE); all & 6568 & 89 \\
Interpolation (SMOTE); accepted & 2337 & 30 \\
\hline
Interpolation (Triangle); all & 6781 & 13 \\
Interpolation (Triangle); accepted & 5620 & 6 \\
\hline
Interpolation (Pairs); all & 16\,508 & 14 \\
Interpolation (Pairs); accepted & 7569 & 10 \\
\hline
Hypersphere; all & 19\,603 & 2068 \\
Hypersphere; accepted & 15\,126 & 1545 \\
\hline
\end{tabular}
\end{table}

\subsection*{\sffamily \large Filtering of candidates for Protein C inhibitors}

Among the known 103 active Protein C ligands, 
39 pass the Lipinski filter, 
15 -- the Ghose filter,
7 -- the Veber filter,
6 -- the Egan filter,
and 9 pass the Muegge filter. 
Only one ligand passes all the five filters.
As the Lipinski filter yields the coarsest sieve, it is interesting to combine it with other filters pairwise.
It appears that 22 known inhibitors pass the Lipinski and at least one of the remaining four filters.

The Ghose filter appears the most strict for the 1571 generated Protein C inhibitor candidates: only one molecule -- from the hypersphere search -- passes it.
As with known inhibitors, the Lipinski filter is the loosest one, with 936 molecules passing it. 
118 generated compounds pass the Veber filter, 101 pass the Egan filter, and 156 -- the Muegge filter.
The Lipinski filter plus at least one more filter approve 218 molecules, while all the filters except the Ghose one are passed by 40 generated molecules.

\begin{figure}[t]
\begin{center}
\includegraphics[scale=0.3]{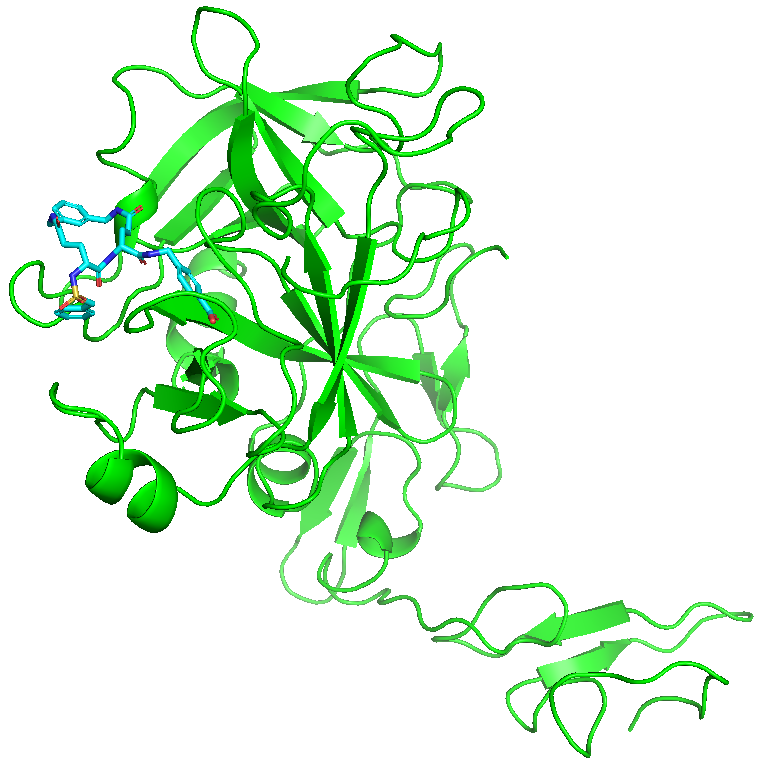}
\caption{Molecular docking performed for one of generated coagulant candidates to Protein C. Protein chains are shown in green, and the molecule is shown in cyan (carbons) / blue (nitrogens) / red (oxygens) / yellow (sulfur) palette.}
\label{fig:protc_docking}
\end{center}
\end{figure}

Another parameter to rank the generated compounds is the Synthetic Accessibility Score (SAS).~\cite{Ertl2009}
The generated Protein C inhibitor candidates sieved through the Ghose, Veber, or Egan filters have SAS $>4$.
The lowest SAS among the generated Protein C candidates is 3.86 and belongs to the hypersphere search.
The lowest SAS values from the closest pairs are 4.17 and 4.3; the respective molecules occur also in the hypersphere and SMOTE sets.
The lowest SAS from the strongest triangle is 6.28; this candidate is common to all the four sets: it is also found in the hypersphere, SMOTE, and closest pairs sets.
As an example, we docked this molecule to the activated form of Protein~C (protein coordinates are taken from the 1AUT structure from Protein Data Bank) using the DiffDock utility.~\cite{Corso2022,DiffDock:www}
The resulting configuration top-ranked is demonstrated in Fig.~\ref{fig:protc_docking}.
The structure of this molecule is shown in Table~\ref{tab:filtered_results} as \#5, where we show a few examples of generated candidates for Protein C inhibitors.
The table includes canonical SMILES, 2D molecular structures, SAS, adherence to the filters, minimal distances in the embedding space to the known strong Protein C inhibitors ($\min d_{\rm P}$) and to the known strong Thrombin inhibitors ($\min d_{\rm T}$). 
Additionally, the approaches generating the respective compound are indicated.
The table is aimed at grasping various kinds of generated ligands, so it includes not only those with the lowest SAS values -- that mostly originate from the hypersphere approach -- but also those obtained through SMOTE, the closest pairs, and the strongest triangle methods. 
Both compounds passing the filters and failing them are included for the sake of wider representation.

\def\imgscale{0.22}

\setlength{\LTcapwidth}{\textwidth}

\begin{longtable}{rp{0.6\textwidth}p{0.3\textwidth}}

\kill
\caption{A selection of generated compounds being candidates for Protein C inhibitors. The listed molecules do not pass the Veber, Egan, and Muegge filters.}\label{tab:filtered_results}
\\[-3ex]
\hline
\# & \qquad\qquad SMILES and information & 
\qquad\qquad Image \\
\hline\endfirsthead

\caption{(\textit{continued from the previous page})}
\\[-3ex]
\hline
\# & \qquad\qquad SMILES and information & 
\qquad\qquad Image \\
\hline\endhead

\noalign{\smallskip}\hline
\endfoot

 1. & \multicolumn{2}{>{\raggedright}p{0.9\textwidth}}{\footnotesize O=C(O)C1=CC=C(CNC(=O)C(CC2=CC=C(O)C=C2)NC(=O)C(CC2(C)CC2)NCC(=O)O)C=C1} \\*
   & SAS = 3.86, Lipinski: Yes; Ghose: No\newline
     $d_{\rm P} = 0.143$; \quad $d_{\rm T} = 0.699$\newline
     Generated by Hypersphere & 
   \raisebox{-0.85\height}{\includegraphics[scale=\imgscale]{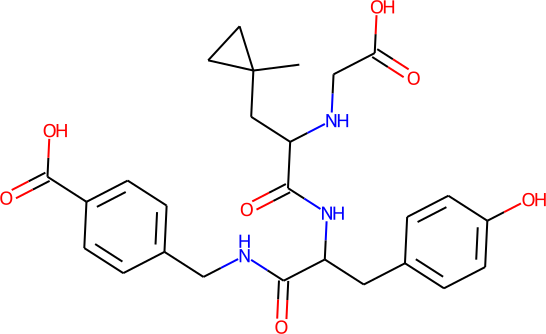}} \\
  2. & \multicolumn{2}{>{\raggedright}p{0.9\textwidth}}{\footnotesize O=C(O)C1=CC=C(CNC(=O)C(CC2=CC=C(O)C=C2)NC(=O)C(CC2CCNCC2)NCC(=O)O)C=C1} \\*
   & SAS = 4.05, Lipinski: No; Ghose: No\newline
     $d_{\rm P} = 0.129$; \quad $d_{\rm T} = 0.709$\newline
     Generated by Hypersphere & 
   \raisebox{-0.85\height}{\includegraphics[scale=\imgscale]{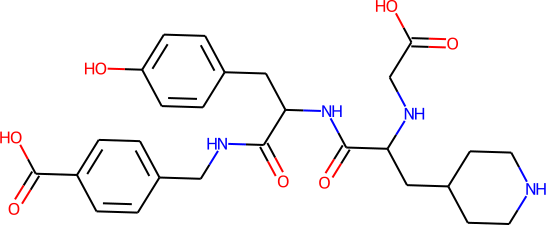}} \\
   3. & \multicolumn{2}{>{\raggedright}p{0.9\textwidth}}{\footnotesize O=C(OC1=CCC1NC(=O)C(CC1CCC=C1)=CC(=O)O)C(=O)NCC1=CC=C(C(=N)N)C=C1} \\*
   & SAS = 4.57, Lipinski: Yes; Ghose: Yes\newline
     $d_{\rm P} = 0.205$; \quad $d_{\rm T} = 0.660$\newline
     Generated by Hypersphere & 
   \raisebox{-0.85\height}{\includegraphics[scale=\imgscale]{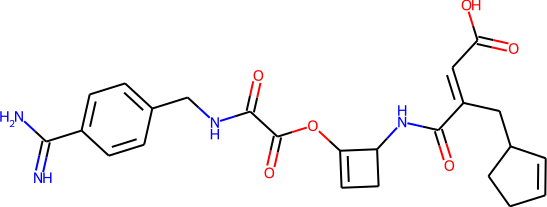}} \\
   4. & \multicolumn{2}{>{\raggedright}p{0.9\textwidth}}{\footnotesize CC(C)CC(=O)NC1=CC(N2CC(C)(C3=CC=CC=C3)C3=CC(C(=N)N)=CC=C3N2)=CC(C2=CC= C(C(N)=O)C=C2C(=O)O)=C1} \\*
   & SAS = 5.13, Lipinski: No; Ghose: No\newline
     $d_{\rm P} = 0.074$; \quad $d_{\rm T} = 0.882$\newline
     Generated by Interpolation (SMOTE) & 
   \raisebox{-0.85\height}{\includegraphics[scale=\imgscale]{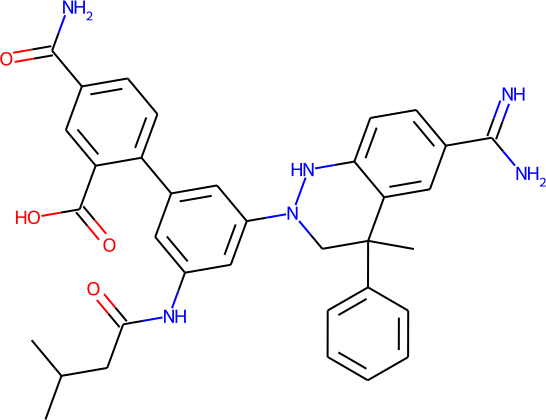}} \\
 5. & \multicolumn{2}{>{\raggedright}p{0.9\textwidth}}{\footnotesize O=C(O)C1=CC=C(CNC(=O)C2CCC(=O)NCC3=CC=CC(=C3)CNC(=O)CCC(NS(=O)(=O)CC3= CC=CC=C3)C(=O)N2)C=C1} \\*   & SAS = 6.28, Lipinski: No; Ghose: No\newline
     $d_{\rm P} = 0.072$; \quad $d_{\rm T} = 0.783$\newline
     Generated by Hypersphere and Interpolation (SMOTE, Pairs, and Triangle) & 
   \raisebox{-0.85\height}{\includegraphics[scale=\imgscale]{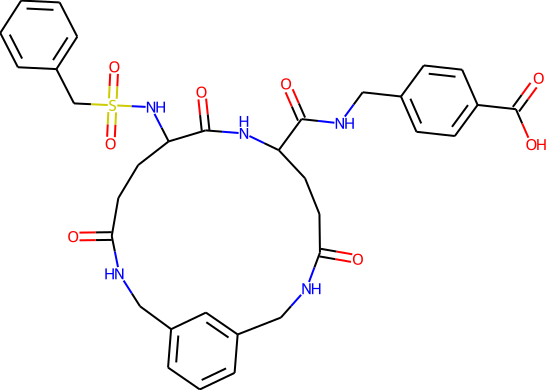}} \\
\end{longtable}

\subsection*{\sffamily \large Generation of candidates for Protein C inhibitors with MegaMolBART}

We also used MegaMolBART to generate new molecules between Protein C inhibitor pairs neighboring in our embedding space, namely, $s=(3, 4)$, (4, 9), (6, 13), (6, 14), and (13, 14); additionally, the hypersphere generation was performed around strong inhibitors $s=2, 3, 4, 6, 7, 8, 9, 10, 11, 13, 14, 15$.
Not all strong inhibitors were engaged as the selection of reference points in MegaMolBART is imposed by their availability in ChEMBL~27.

For the interpolation between inhibitor pairs, 100 molecules were requested for each pair by probing at regularly spaced intervals in the latent space of the MegaMolBART model.
For the generation around a specified inhibitor, up to 60 molecules were requested. 
An integer scaled sampling radius parameter $R_{\rm s}$ was set to the lowest value of 1.
At larger radii, no molecules were yielded for the majority of reference inhibitors.
Exceptions are represented by $s = 6, 8, 9, 14, 15$ with 2--3 generated molecules at $R_{\rm s} = 2$.
The generated set was further filtered for unique SMILES.
Some generated compounds from both our approach and MegaMolBART are compared in Tables~\ref{tab:compare-gen-img2}--\ref{tab:compare-gen-img3}.
\def\imgscale{0.19}
\def\imgscalebig{0.25}
\begin{table}[p]
\caption{Comparison of molecules generated by interpolation between strong Protein C inhibitors.}
\label{tab:compare-gen-img2}

\smallskip
    \centering
\begin{tabular}{m{0.49\textwidth}|m{0.49\textwidth}}
\hline\noalign{\smallskip}
\multicolumn{2}{c}{
\includegraphics[scale=\imgscale]{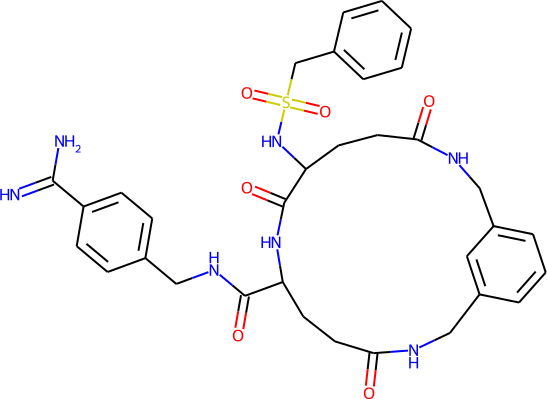}
\raisebox{8ex}{$\longleftarrow$ Reference ligands $s=3,4$ $\longrightarrow$}
\includegraphics[scale=\imgscale]{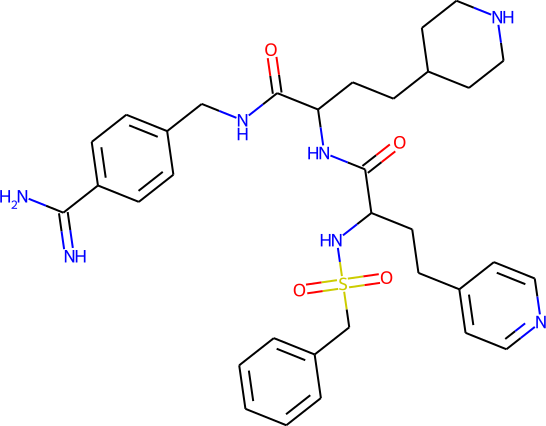}
} \\
\hline
 
\multicolumn{1}{c}{Our approach} &
\multicolumn{1}{c|}{MegaMolBART}

\\
\hline\noalign{\smallskip}

\includegraphics[scale=\imgscale]{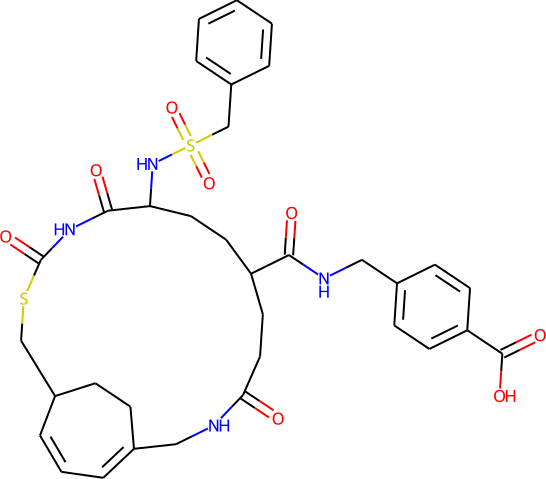}
\quad
\includegraphics[scale=\imgscale]{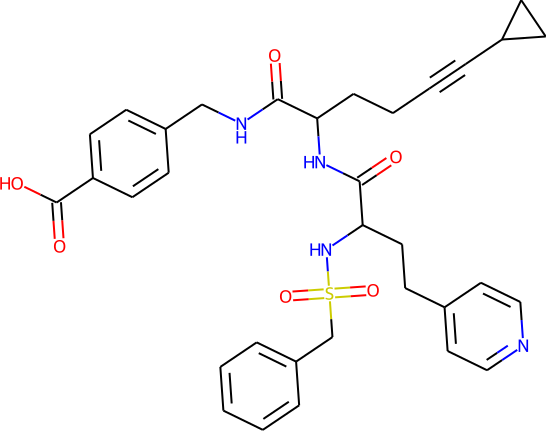} 

\includegraphics[scale=\imgscale]{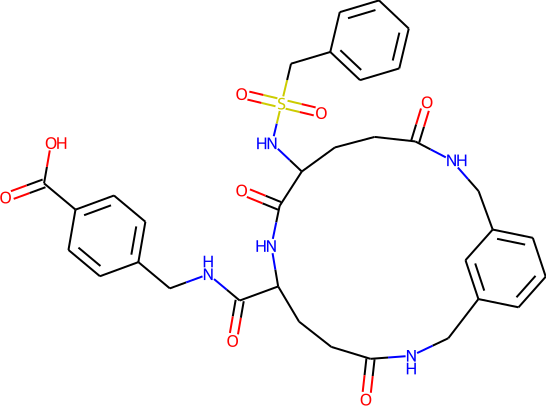}
\quad
\includegraphics[scale=\imgscale]{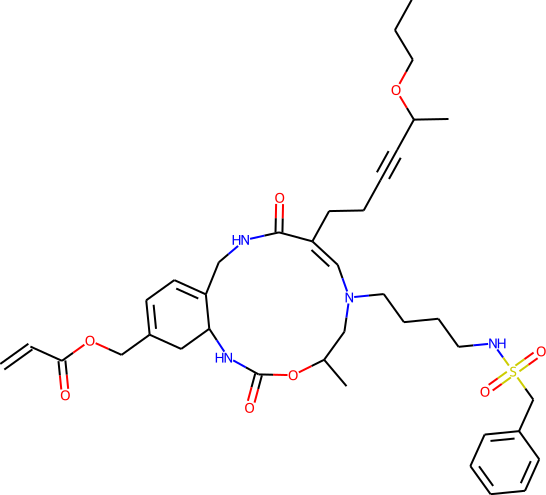}

&
\includegraphics[scale=\imgscale]{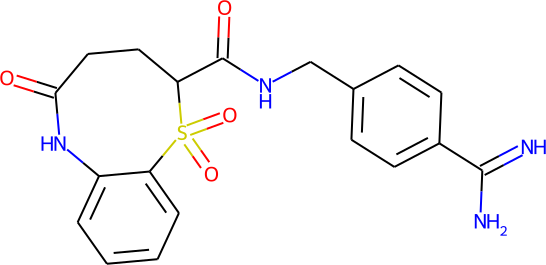}
\ 
\includegraphics[scale=\imgscale]{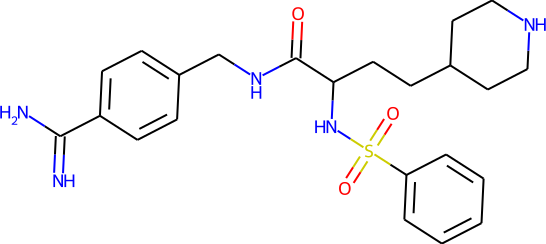} 
\ 
\includegraphics[scale=\imgscale]{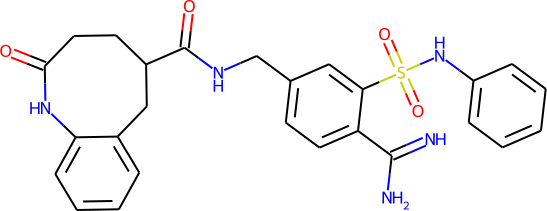}
\ 
\includegraphics[scale=\imgscale]{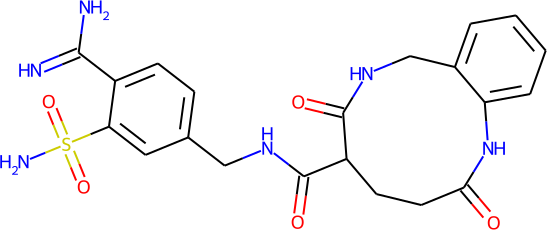}
\ 
\includegraphics[scale=\imgscale]{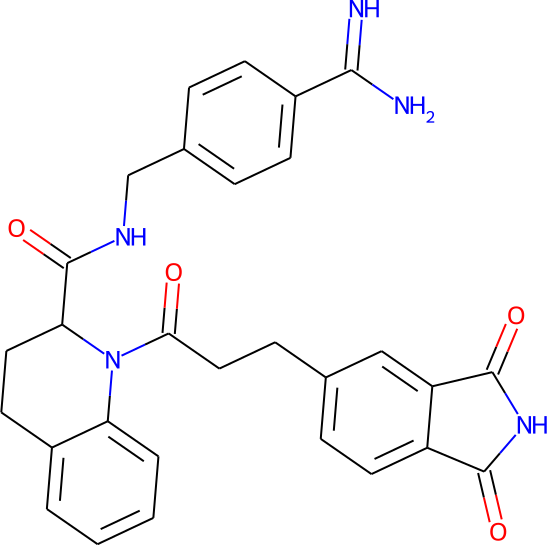}
\ 
\includegraphics[scale=\imgscale]{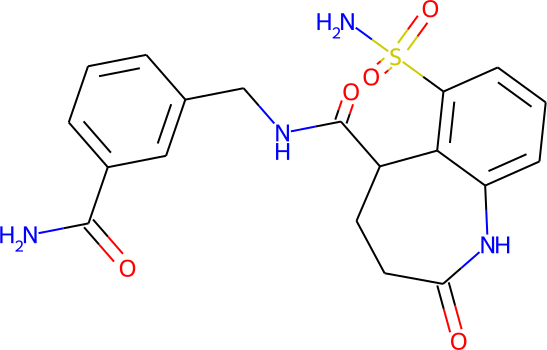}
\\
\noalign{\smallskip}\hline

\hline\noalign{\smallskip}
\multicolumn{2}{c}{
\includegraphics[scale=\imgscale]{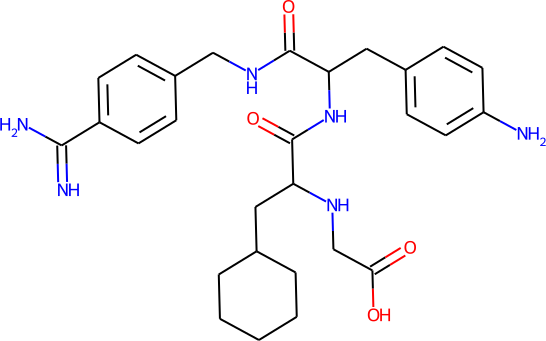}
\raisebox{8ex}{$\longleftarrow$ Reference ligands $s=6,14$ $\longrightarrow$}
\includegraphics[scale=\imgscale]{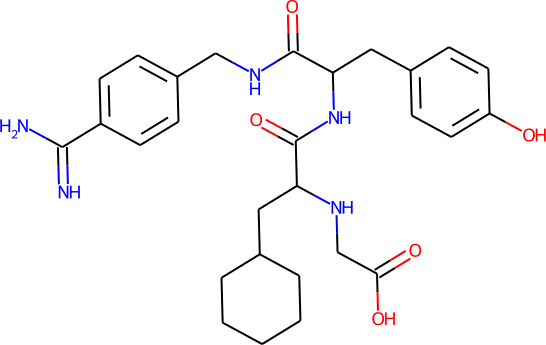}
} \\
\hline
\multicolumn{1}{c}{Our approach}
&
\multicolumn{1}{c|}{MegaMolBART}\\
\hline\noalign{\smallskip}

\includegraphics[scale=\imgscale]{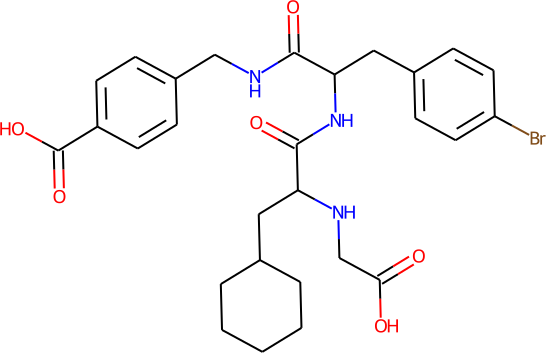}
\quad
\includegraphics[scale=\imgscale]{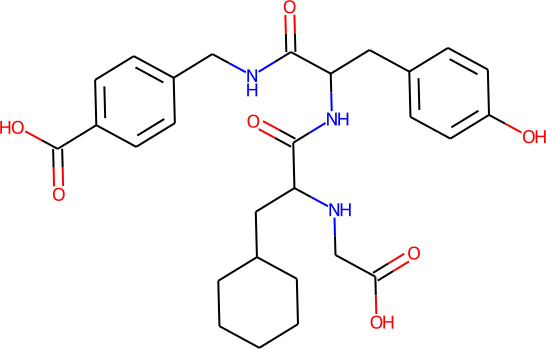}

\hspace*{6em}
\includegraphics[scale=\imgscale]{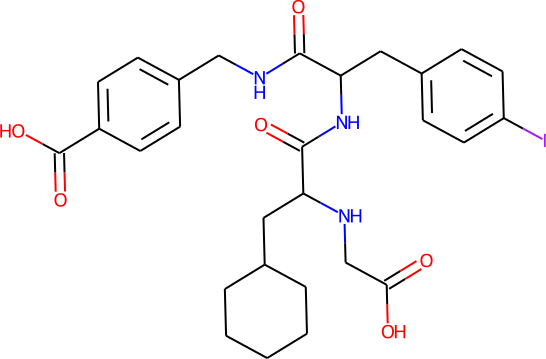}

& 
\hspace*{4em}\raisebox{2em}
{%
\includegraphics[scale=\imgscalebig]{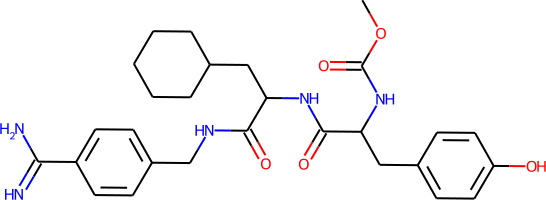}}

\hspace*{4em}\raisebox{-2em}
{%
\includegraphics[scale=\imgscalebig]{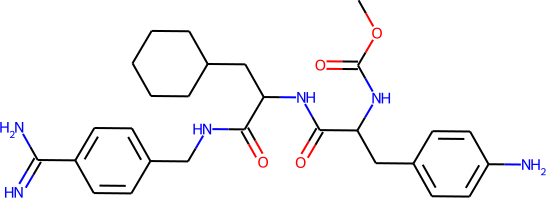}} 
\\
\noalign{\smallskip}\hline
\end{tabular}
\end{table}

\begin{table}[t]
\caption{Comparison of molecules generated around the strong Protein C inhibitor $s=3$.}
\label{tab:compare-gen-img3}

\smallskip
    \centering
\begin{tabular}{b{0.49\textwidth}|b{0.49\textwidth}}
\hline\noalign{\smallskip}
\multicolumn{2}{c}{
\raisebox{8ex}{Reference ligand:\qquad}
\includegraphics[scale=\imgscale]{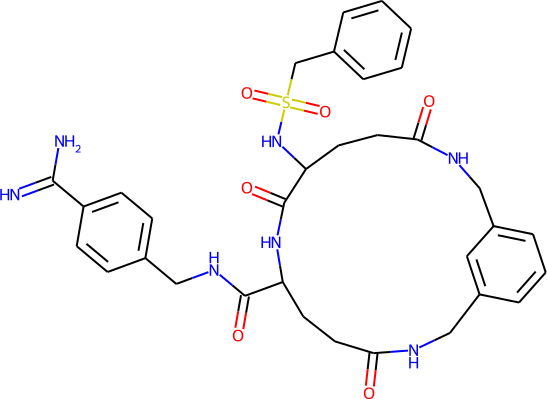}
} \\

\hline
\multicolumn{1}{c}{Our approach}
&
\multicolumn{1}{c|}{MegaMolBART}
\\
\hline\noalign{\smallskip}

\includegraphics[scale=\imgscale]{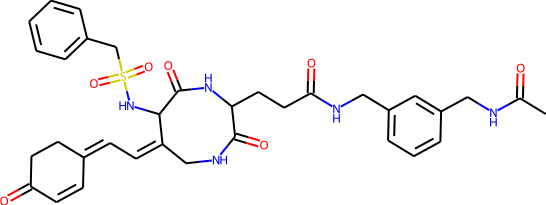} 
\ 
\includegraphics[scale=\imgscale]{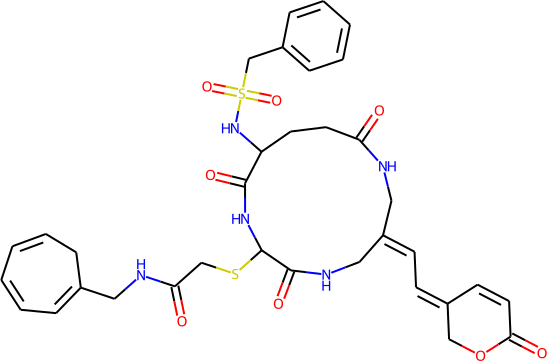}
&
\includegraphics[scale=\imgscale]
{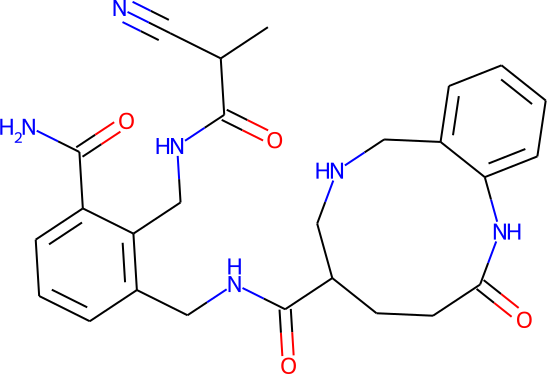}
\ 
\includegraphics[scale=\imgscale]
{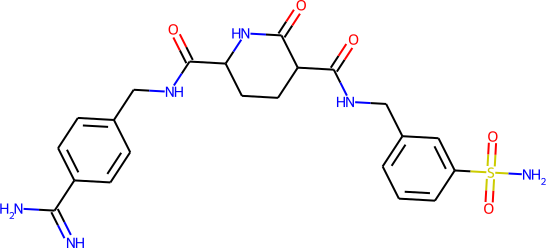}
\\
\includegraphics[scale=\imgscale]{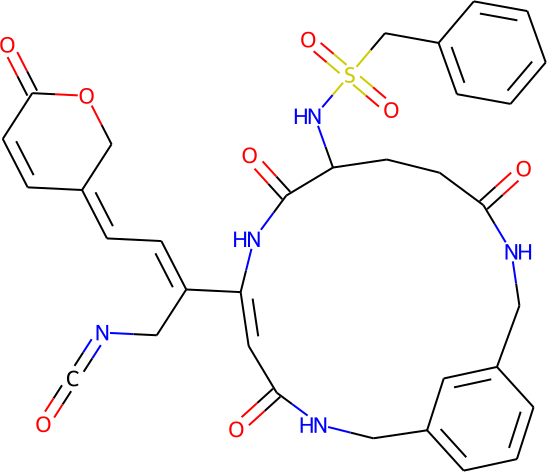}
\ 
\includegraphics[scale=\imgscale]{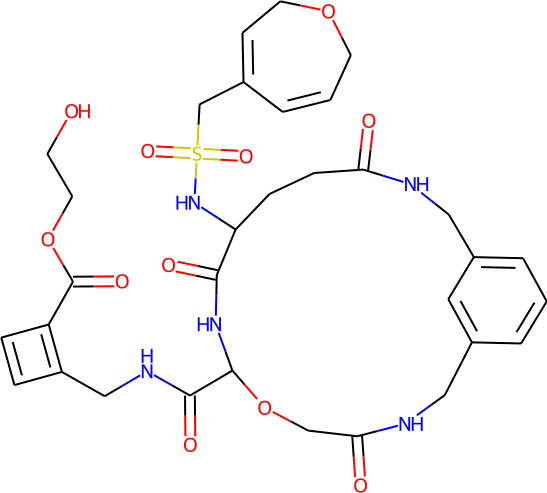}
&
\includegraphics[scale=\imgscale]
{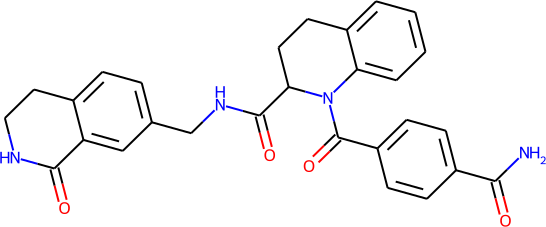}
\ 
\includegraphics[scale=\imgscale]
{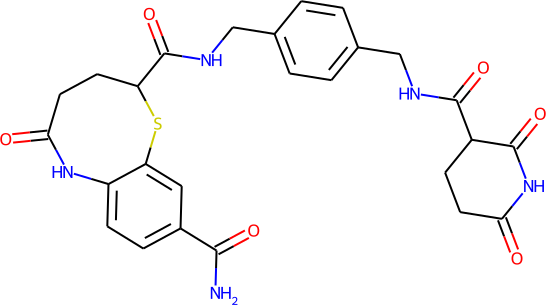}
\\
\includegraphics[scale=\imgscale]{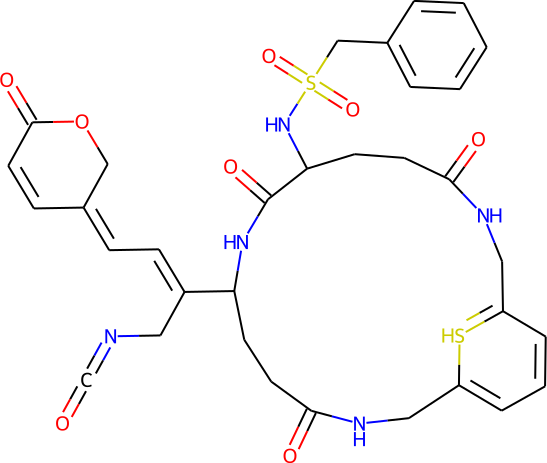}
\ 
\includegraphics[scale=\imgscale]{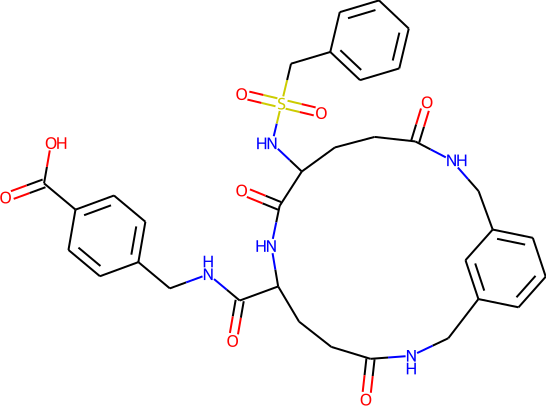}
&
\includegraphics[scale=\imgscale]
{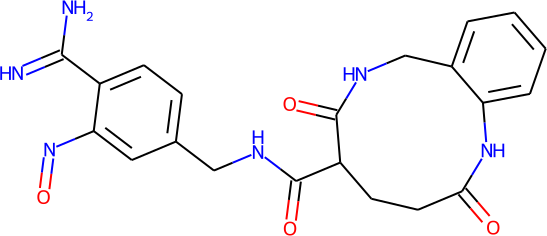}
\ 
\includegraphics[scale=\imgscale]{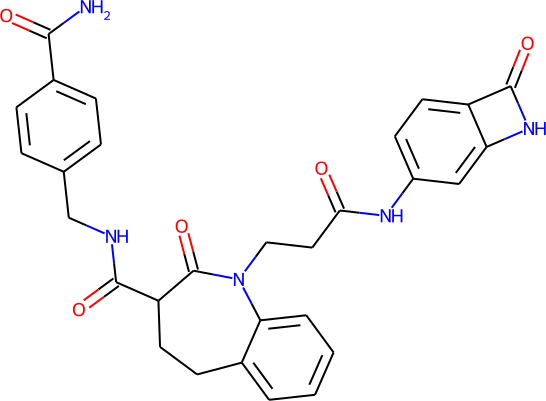} 
\\
\noalign{\smallskip}
\hline
\end{tabular}
\end{table}

From Tables~\ref{tab:compare-gen-img2}--\ref{tab:compare-gen-img3}, one can conclude that our approach tends to preserve structures of reference compounds.
This is best seen in Table~\ref{tab:compare-gen-img3} containing molecules generated around a given reference inhibitor.
On the other hand, one can notice an appearance of cyclopropane rings in \ \raisebox{-0.2ex}{\includegraphics[scale=0.12]{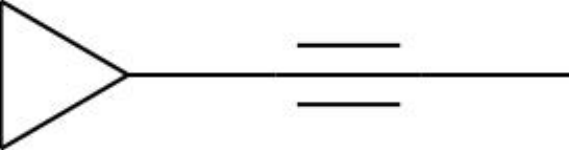}} \ corresponding to a \texttt{CC\#CC1CC1} fragment and cyclobutadiene rings in \ \raisebox{-1ex}{\includegraphics[scale=0.1]{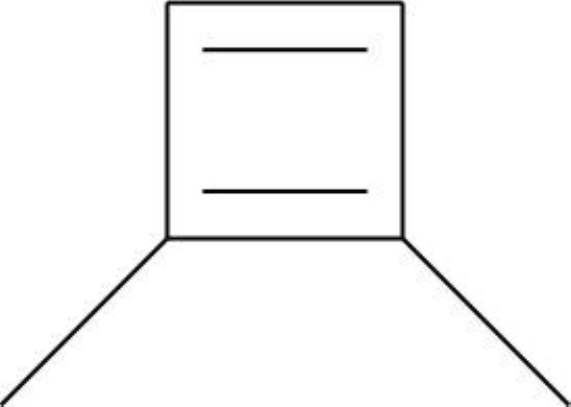}} \ corresponding to a \texttt{CC1=C(C)C=C1} fragment.
These are not found in the original reference inhibitors.
Another interesting observation is that the number of molecules generated by MegaMolBART between close inhibitors exceeds that from our approach in most cases.
For the pair $s=3,4$ we have 14 molecules generated by MegaMolBART \textit{vs} 6 molecules in our approach.
Similarly, it is 6 \textit{vs} 3 for $s = 4,9$, 14 \textit{vs} 3 for $s = 6,13$, and 12 \textit{vs} 2 for $s = 13,14$.

However, as shown in Table~\ref{tab:compare-gen-img2}, there are only two distinct SMILES generated by MegaMolBART between reference ligands $s=6,14$.
The number did not change even upon requesting to generate 1000 molecules. 
For the same reference ligands, our approach yields three distinct SMILES.
Of the analyzed pairs, only $s=6,14$ seem to be closely located in both latent spaces.
Discrepancies in the numbers by the two approaches are rooted in differences between our embedding space and the space of the MegaMolBART model.
The proximity of reference points in one space does not necessarily mean they should be close in another space, however, a thorough work with MegaMolBART source code falls beyond the scope of this study.

All the experiments with MegaMolBART
were run on an NVIDIA A100 GPU.

\section*{\sffamily \Large CONCLUSIONS}
\label{sec:Conclusions}

In this study, we tackled the problem of the generation of novel chemical compounds and for this purpose we chose a specific case: generation of coagulants.
One of the ways to enhance blood coagulation is inhibition of proteins that provide anticoagulatory function, thus, shifting the balance towards coagulation.
In a simplified setup, when there are many known inhibitors of a target protein, one can build a data-driven prediction model, screen a given list of compounds to filter, and rank them by inhibition strength.
In particular, a candidate target to inhibit its action is Protein C.
A problem formulation in the case of Protein C reads: there is a limited number of its known inhibitors, hence, an alternative is needed.
For this purpose, we utilized an autoencoder -- one of machine learning concepts -- which allows for training on millions of arbitrary molecules.
The overarching majority of these molecules have no inhibitory action towards a target protein but certainly contribute to the knowledge of chemical rules about the mutual arrangement of atoms and bonds.
Autoencoders convert chemical structures into their compressed numerical representations, still obeying the above rules.
The set of compressed representations is further considered as a ``map of chemical compounds''.
Hence, one can use known inhibitors (even if there is a small number of them) as reference points on the map and inspect their vicinity for other compounds.
These new compounds will differ from original inhibitors by small structural variations and might reveal inhibitor function too.
We validated different exploration techniques (traversing over a hypersphere around reference inhibitors, interpolation between two close inhibitors, or population within a triangle of three close inhibitors) and discussed their comparison.
Additionally, the search of coagulants is complemented by the generation of anticoagulants (inhibitors of Thrombin).
A complex look at the coagulant/anticoagulant generation implies the analysis of mutual locations of known inhibitors to define typical ``distances'', not to overgenerate coagulant candidates in the vicinity of anticoagulants and vice versa.
The autoencoder approach is compared to another machine learning-based solution -- MegaMolBART.
MegaMolBART originates from the family of so-called large language models and works on slightly different principles.
It also offers generating candidate compounds either within a hypersphere or by interpolation between reference inhibitors.
Thus, it is interesting to spot the behavior peculiarities of the two generative approaches -- such differences only enrich the palette of suggested candidates.
It is worth emphasizing that the considered Protein C case is an example scenario -- the approach can be scaled to other proteins as well.


\end{document}